\documentclass[preprint,12pt,3p]{elsarticle}

\usepackage[utf8]{inputenc}
\usepackage{xcolor}
\usepackage{hyperref}
\hypersetup{
    colorlinks=true,
    citecolor=blue,
    linkcolor=blue,
    filecolor=magenta,      
    urlcolor=blue,
}
\usepackage{graphicx}
\usepackage{amssymb}
 \usepackage{amsthm}

\usepackage{lineno}

\usepackage{booktabs}

\usepackage{color, colortbl}

\biboptions{comma,square}

\journal{Epilepsy \& Behavior}

\begin{document}

\begin{frontmatter}

%\title{Increased activity on \textit{Facebook} may help predict Sudden Unexpected Death in Epilepsy: A Preliminary Study}

\title{Small Cohort of Epilepsy Patients Showed Increased Activity on \textit{Facebook} before Sudden Unexpected Death}

%% Authors
\author[CSBC]{Ian B. Wood\corref{con}}
\author[IGC,CAPES,CSBC]{Rion Brattig Correia\corref{con}}
\author[NURSING]{Wendy R. Miller\fnref{cor1}}
\author[SUNY,CSBC,IGC]{Luis M. Rocha\fnref{cor2}}
%% Correspondence (Emails)
\fntext[cor1]{wrtruebl@iu.edu}
\fntext[cor2]{rocha@binghamton.edu}
\cortext[con]{authors contributed equally}
%% Affiliation
\address[CSBC]{Center for Social and Biomedical Complexity, Luddy School of Informatics, Computing \& Engineering, Indiana University, Bloomington, IN 47408 USA}
\address[IGC]{Instituto Gulbenkian de Ciência, Oeiras 2780-156, Portugal}
\address[CAPES]{CAPES Foundation, Ministry of Education of Brazil, Brasília, DF, Brazil}
\address[NURSING]{School of Nursing, Indiana University, Indianapolis, IN 46202 USA}
\address[SUNY]{Department of Systems Science and Industrial Engineering, Binghamton University, Binghamton, NY 13902}

%
% Abstract
%
\begin{abstract}
Sudden Unexpected Death in Epilepsy (SUDEP) remains a leading cause of death in people with epilepsy.
Despite the constant risk for patients and bereavement to family members, to date the physiological mechanisms of SUDEP remain unknown.
Here we explore the potential to identify putative predictive signals of SUDEP from online digital behavioral data using text and sentiment analysis.
Specifically, we analyze \textit{Facebook} timelines of six epilepsy patients deceased due to SUDEP, donated by surviving family members.
We find preliminary evidence for behavioral changes detectable by text and sentiment analysis tools.
Namely, in the months preceding their SUDEP event patient social media timelines show:
i) increase in verbosity;
ii) increased use of functional words;
and iii) sentiment shifts as measured by different sentiment analysis tools.
Combined, these results suggest that social media engagement, as well as its sentiment, may serve as possible early-warning signals for SUDEP in people with epilepsy.
While the small sample of patient timelines analyzed in this study prevents generalization, our preliminary investigation demonstrates the potential of social media data as complementary data in larger studies of SUDEP and epilepsy.
\end{abstract}


\end{frontmatter}

%\linenumbers

\section{Introduction}
\label{ch:introduction}

Sudden Unexpected Death in Epilepsy (SUDEP) remains a leading cause of death for people with epilepsy (PWE), and includes all epilepsy-related deaths not due to trauma, drowning, status epilepticus, or other identifiable causes.
The incidence of SUDEP is about 0.35 cases per 1,000 person-years \cite{bagnall2017genetic}.
While research into the physiological mechanisms underlying SUDEP continue to be thoroughly studied, and new SUDEP-related guidelines for clinicians treating PWE have been published in order to minimize SUDEP risk, SUDEP incidence remains steady \cite{Miller:2014:SUDEP, harden2017practice}.
To date, the most espoused preventive strategy for SUDEP remains seizure control via appropriate self-management \cite{smithson2014sudden}, and especially medication adherence, since a clear risk factor for SUDEP is a higher frequency of seizures \cite{CDC}.
While these risk factors have been disseminated broadly, including to the public, SUDEP remains a leading cause of death for PWE, leading organizations such as The Institute of Medicine, American Epilepsy Society, and Epilepsy Foundation to call for increased study into SUDEP.

Apart from research related to the ways in which providers, patients, and their families discuss SUDEP \cite{Miller:2014:SUDEP, stevenson2014knowing}, very little behavioral research has been conducted to reveal potential behavioral or social attributes that may precede SUDEP.
Should such specific attributes exist, they would provide an area of preventive intervention for SUDEP.
In this study, we utilize digital behavioral data and investigate its potential for uncovering behavioral signatures preceding SUDEP that could be leveraged as early-warning signals to inform self-management interventions in PWE.
As patients are known to not fully recall important events or even display recognizable behavior change during clinical consultations, digital behavioral data, such as social media data, can offer a complementary view of patient behavior of clinical significance \cite{Correia:2020}.
Specifically, we use text and sentiment analysis to evaluate temporal changes in emotional states and communication patterns of the subjects in the study.
The methodology gives us the unique opportunity to examine longitudinally the emotional states of a cohort of PWE with a known outcome of SUDEP.
Our preliminary results show that social media may reveal behavioral experiences leading up to SUDEP, and thus guide areas for SUDEP-preventing interventions.
This study also demonstrates the successful use of alternative, real-world data sources in studying SUDEP \cite{Correia:2020,Ramagopalan:2020}.

Psychological stress is known to increase the risk of certain diseases, like the common cold \cite{cohen1991psychological}.
Directly related to PWE, stress and major life events are known to increase the risk of seizures, which in turn can increase the risk of SUDEP \cite{McConnell:1995, McKee:2017}. However, direct physiological measurements of stress involves expensive and invasive tools. 
A compelling alternative is to measure stress and other cognitive states indirectly in self-reported digital behavioral data, such as in social media posting on \textit{Facebook}.
This is one of the focuses of the interdisciplinary field of \textit{affective computing}, which has developed methods to measure human emotion (including stress) via linguistic and other computer-based features, such as keystroke dynamics \cite{zhai2006stress}.
For instance, Pennebaker \cite{pennebaker1993putting} found a correspondence between textual features and physiological signals of stress.
Similarly, Vizer, Zhou, \& Sears \cite{vizer2009automated} found that increased lexical complexity (diversity of words) tends to correspond with increased physical or cognitive stress.
However, such studies are often conducted in controlled laboratory conditions, asking participants to write essays with particular prompts.
This is not the case with social media, where users write posts spontaneously without being prompted in laboratory settings. 
Our assumption is that stress and other mood states influence whether and how a social media post is written, and can thus be measured via textual analysis of those posts. 
A substantial body of literature already reports that social media data enables quantitative measurement and prediction of various behavioral processes of biomedical relevance, i.e. a real-world data source to study ``humans as their own model organism'' \cite{Correia:2020}.
Indeed, social media data has already been shown to be useful, alone or in combination with other data sources, for a variety of other biomedical problems. For instance, data from Twitter and Instagram helps in the detection of health conditions including the spread of flu pandemics \cite{Christakis:2010}, warning signals of drug adverse reactions \cite{Correia:2016}, human reproduction \cite{wood2017human}, and even depression \cite{Choudhury:2013a}. Social media users who self-reported their diagnosis of depression have been shown to exhibit distorted modes of thinking (cognitive distortions) in their writing, an early warning that can lower the burden of this underdiagnosed condition and leading cause of disability worldwide \cite{Bathina:2021}. A long list of successful applications using social media data for biomedical and health-related problems is discussed in our recent review \cite{Correia:2020}.

To infer relevant cognitive states in our cohort of deceased SUDEP subjects we use textual and sentiment analysis of their social media posts.
These methods were originally developed to determine the positive or negative feelings expressed in natural language texts towards specific product ratings, often used for marketing purposes \cite{Pang:2008, Liu:2012}.
However, a number of sentiment analysis tools have been developed from psychological experiments, and can be used to model the emotional states of authors based on their written text \cite{Correia:2020}.
In fact, sentiment analysis has been very useful to track various individual and cohort specific behaviors of relevance to biomedicine, especially mental health \cite{wood2017human, Correia:2020, Bathina:2021}.
Similarly to other domains, these computational methods are likely to be useful to characterize the behavior of SUDEP cohorts, including any possible stress markers hidden in their social media discourse that can be leveraged to inform interventions aimed at improving self-management, a key predictor of epilepsy-related outcomes.
Next, we detail the data gathering, textual methods, and three different sentiment analysis tools we apply to our SUDEP cohort.

\section{Materials and methods}
\label{ch:methods}

%Methods used to recruit families to donate user timelines, and to collect and analyze the \textit{Facebook} data on SUDEP decease individuals, are detailed in this section.

We began by eliciting families from which a member was known to have died of SUDEP. To do so, we advertised our research goals on the bulletin boards of the Epilepsy Foundation website and epilepsy-related \textit{Facebook} groups. We also distributed information about our study to the Epilepsy Foundation's SUDEP Institute, which passed on the information to members of SUDEP bereavement groups within the Institute.
The Epilepsy Foundation website is one of the most popular sites for people with epilepsy.
All procedures were evaluated by the Indiana University Institutional Review Board, who ultimately deemed that the study was exempt/not human subjects research. Family members self-referred to our study via email, and were given information about the study, its goals, and were also informed that participation was voluntary.  
We received about 20 inquiries from families who wanted to donate social media content from their deceased family members.
From these, a majority of users had \textit{Facebook} accounts, and only a few had \textit{Twitter} or \textit{Instagram} accounts.
Due to data availability we decided to focus our analysis solely on \textit{Facebook} timelines.
This yielded a small cohort of $n=12$ \textit{Facebook} timelines (four males and eight females) from which we had timelines to collect data from.
For six subjects we obtained full login information, and for the remaining we had varying viewing access to timeline posts, as listed in Table \ref{tab:subject-demographics}.

Data collection for subjects with login information was conducted through an in-house developed application using \textit{Facebook}'s official application programming interface (API). Family members logged into the deceased \textit{Facebook} account and accessed a specific app webpage. The app then collected all of the subject's timeline posts, including text, meta-data (e.g., date, posting device, etc), and the number of likes, comments and shares.
Similarly, when only viewing access to the subject's timeline was available, family members (or a researcher when family was unable/unavailable) were instructed to scroll the deceased timeline, thus loading all posts, and export the subject's timeline content as an \textsc{html} file.
A script developed in-house was used to process the \textsc{html} file, collecting text, available meta-data, and number of likes, comments, and shares from posts. %, as well as comments from friends.
Importantly, unlike the app-collected timelines that made use of subject's login information, timelines collected via the \textsc{html}-scraping script may not contain all subject posts, as privacy settings putatively put in place by the subject may have blocked the person collecting the data from viewing them in the first place.
In addition, in 2009 \textit{Facebook} made a significant change to their interface: the prompt to the post box changed from ``Update your Status'', followed by ``$<$Subject name$>$ is...'' to ``What's on your mind?''. Naturally, we believe this interface change may elicit a different response from the user. To avoid any possible interface bias in our analysis, we only consider subject posts that occurred after 2009, when the change took place.
All collected data were securely stored within our servers for further analysis. For each subject Table \ref{tab:subject-demographics} lists basic demographic, subject posting time range, and any notable life event discussed by the subject on their \textit{Facebook} timeline in the month preceding their SUDEP, which was manually annotated by the researchers.

The number of posts collected for each subject varies widely, from only 4 posts written by Subject 12, all the way to 2,271 posts written by Subject 2 (see Table \ref{tab:subject-demographics}).
The average number of posts per subject is 726.
In total, we collected and processed 8,717 posts with text that were written after 2009, when considering all 12 subjects.
However, because some subjects had very little number of posts---as is the case of Subject 12---we opted to limit our analysis to subjects with more than 500 posts that contained text and were written after 2009.
In other words, next we only present results on subjects 1-3, 6, 8, and 10, a cohort of $n=6$ subjects.
These subjects are highlighted in Table \ref{tab:subject-demographics}.

\definecolor{gray}{gray}{0.95}

\begin{table}[t]
    \footnotesize
    \centering
    \begin{tabular}{c|l|c|c|r|l|l|l}
    \toprule
    Subj. & Collection & Sex & Age & Posts & Window of posts\textsuperscript{*} & Notable life event before SUDEP \\
    \midrule
    \rowcolor{gray}
    1 & App & F & 23 & 1,410 & 2,526 & New apartment, job, and city \\
    \rowcolor{gray}
    2 & App & M & 20 & 2,271 & 2,157 & Releasing DVD copies of new movie \\
    \rowcolor{gray}
    3 & App & F & 18 & 844 & 2,071 & Lonely as new college freshman \\
    4 & App & F & 24 & 273 & 1,865 & Graduating a Master's program \\
    5 & App & M & 14 & 51 & 911 & Birthday \\
    \rowcolor{gray}
    6 & App & F & 15 & 473 & 843 & Return from Europe trip\\
    7 & FoF & F & 29 & 62 & 2,334 & n/a \\
    \rowcolor{gray}
    8 & Public & F & n/a & 2,201 & 2,315 & n/a \\
    9 & Public & F & n/a & 10 & 52 & Party and writing paper \\
    \rowcolor{gray}
    10 & Friend & M & 24 & 984 & 2,373 & Recent concussion and recovery \\
    11 & FoF & M & 28 & 134 & 1,524 & Hospitalization \\
    12 & Friend & F & 16 & 4 & 413 & Braces Removed \\
    \bottomrule
    \end{tabular}
    \caption{
        Demographics and data collection details for study subjects.
        Six subject timeline posts were collected via a custom-built app accessed using subject's login and password information.
        Six subject timelines were collected via \textsc{html} scraping of pages as visible to the public, to \textit{Facebook} friends, or to friends of friends (FoF), as noted.
        The number of posts column tallied only posts with written text after 2009 (due to a significant \textit{Facebook} interface change).
        \textsuperscript{*} Column ``window of posts'' denote the number of days between a subject's first and last post.
    }
    \label{tab:subject-demographics}
\end{table}

Textual content of individual posts were processed using the dictionaries of three sentiment analysis tools: \textit{Affective Norms for English Words} (ANEW) \cite{Bradley:1999}, \textit{Valence Aware Dictionary for sEntiment Reasoning} (VADER) \cite{Hutto:2014}, and \textit{Linguistic Inquiry and Word Count} (LIWC) \cite{Tausczik:2010}.
These three tools are widely used in the sentiment analysis literature.
In fact, VADER and LIWC were consistently among the best tools for 3-class polarity classification (negative, neutral, or positive emotion) across a number of corpora in a benchmark comparison study \cite{Ribeiro:2016:SentiBench}.

Dictionaries were used to match against single words in subject posts. Matched words were then scored over several sentiment and textual dimensions per post. 
For instance, ANEW includes ratings from 1 to 9 in a dictionary of 1,034 words along three dimensions: \emph{valence}, from unhappy to happy; \emph{arousal} from calm to excited; and \emph{dominance} from controlled to in-control. These ratings were originally collected from surveys given to undergraduates in a psychology class using a 9-point Likert-like scale \cite{Bradley:1999}. We used ANEW to find the mean sentiment along these three dimensions for each post by averaging the sentiments of each word, while neglecting words absent from the dictionary.
VADER \cite{Hutto:2014} is a tool for measuring the intensity of positive or negative affect through lexical scores modified by syntactical rules, and is readily available as part of the Natural Language Toolkit for python \cite{NLTK:2009}. In addition to dictionary-based sentiment scores, VADER looks at nearby words and modifies sentiment scores based on 5 simple rules: the presence of exclamations, capitalization, adverbs, negations, and contrasting conjunctions. Using this tool, we computed normalized scores describing the intensity of positive, neutral, and negative emotion present in each subject post.
LIWC (pronounced Luke) is the third dictionary-based tool used. It was developed with a well-documented procedure of consistent categorization between a majority of human judges. The latest version of the software, \textsc{LIWC2015}, has dictionaries containing nearly 6,400 words and evaluates text across nearly 90 linguistic and sentiment variables, including summary variables, pronouns, articles, cognitive processes, time focus, personal concerns, and informal language categories \cite{Tausczik:2010, Pennebaker:2015}.

\section{Results}
\label{ch:results}

\begin{figure}[h!]
    \centering
    \includegraphics[width=.3\textwidth]{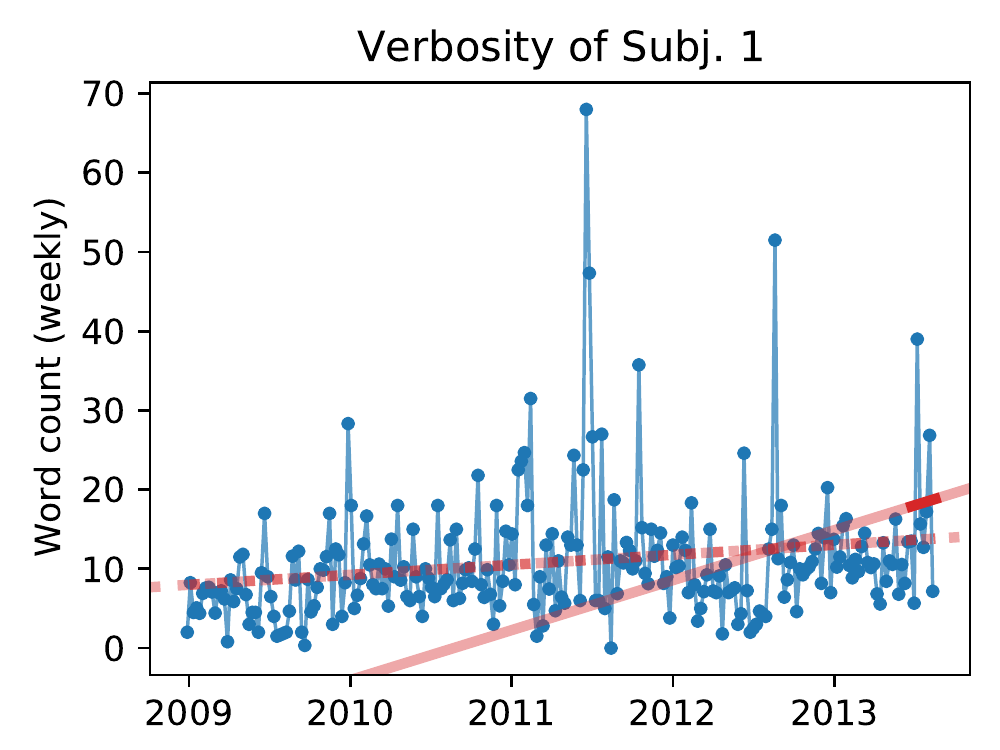}
    \includegraphics[width=.3\textwidth]{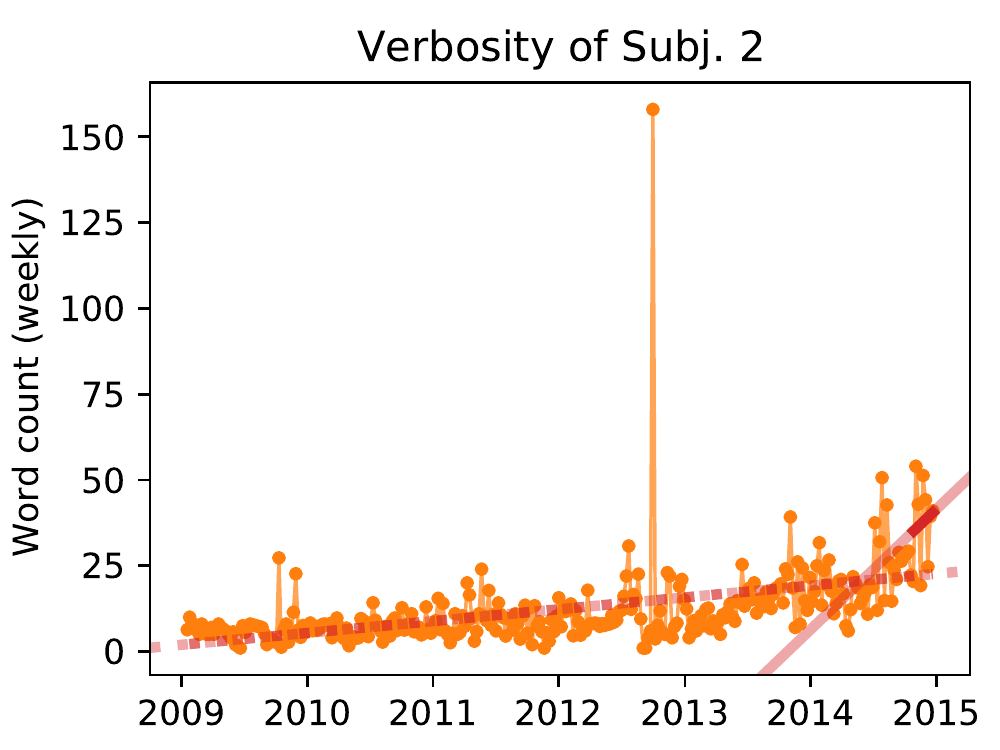}
    \includegraphics[width=.3\textwidth]{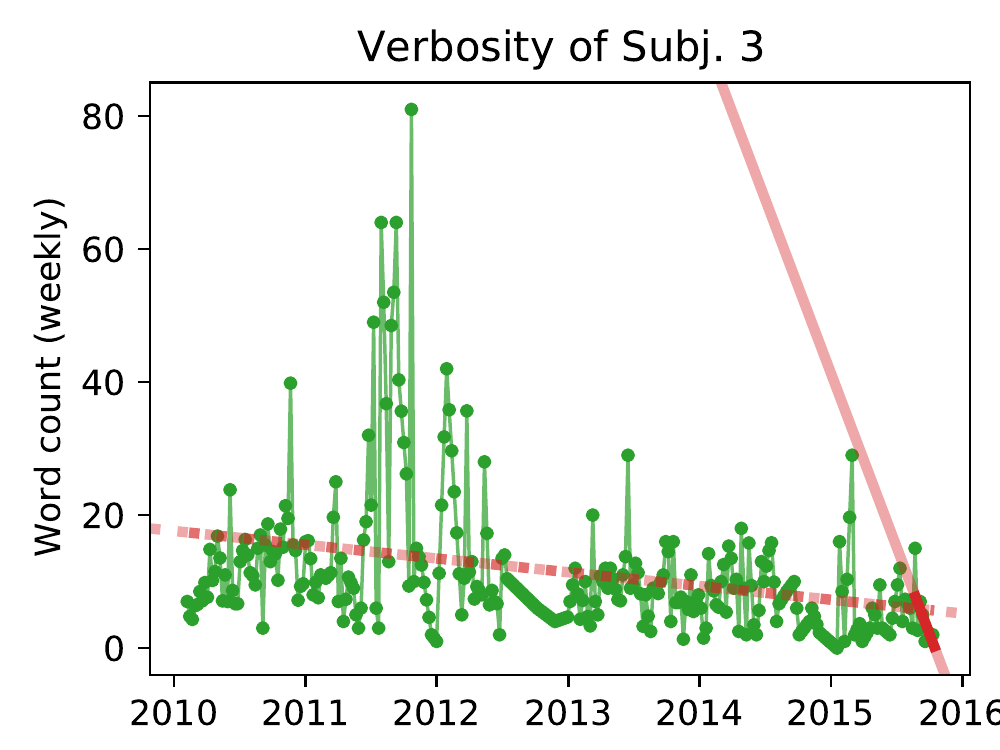}
    \includegraphics[width=.3\textwidth]{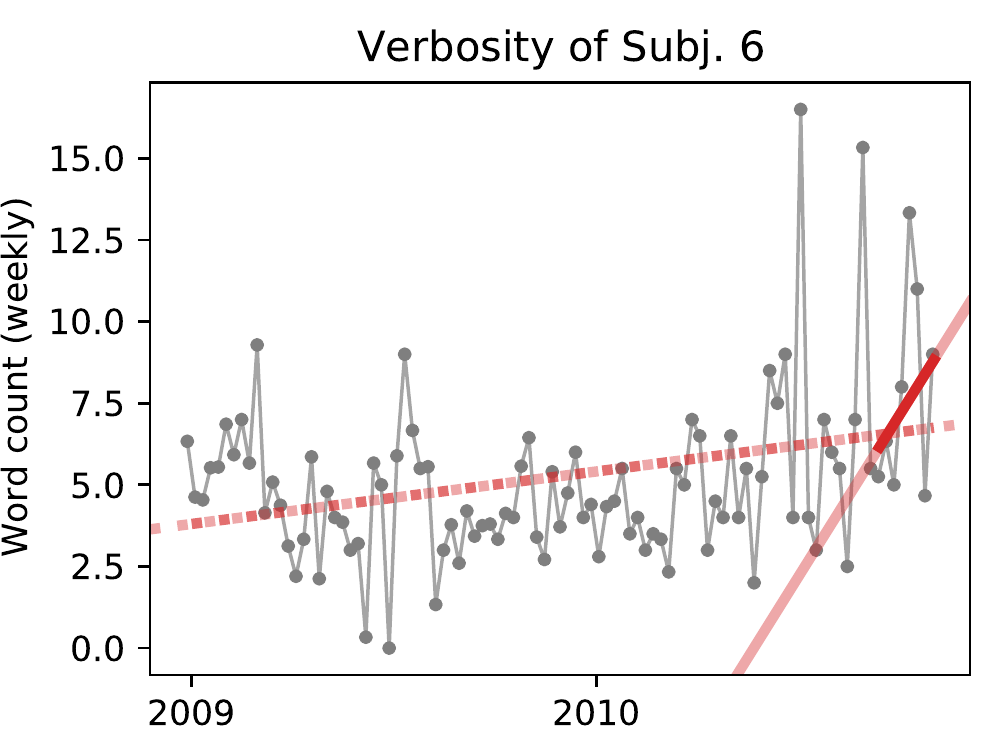}
    \includegraphics[width=.3\textwidth]{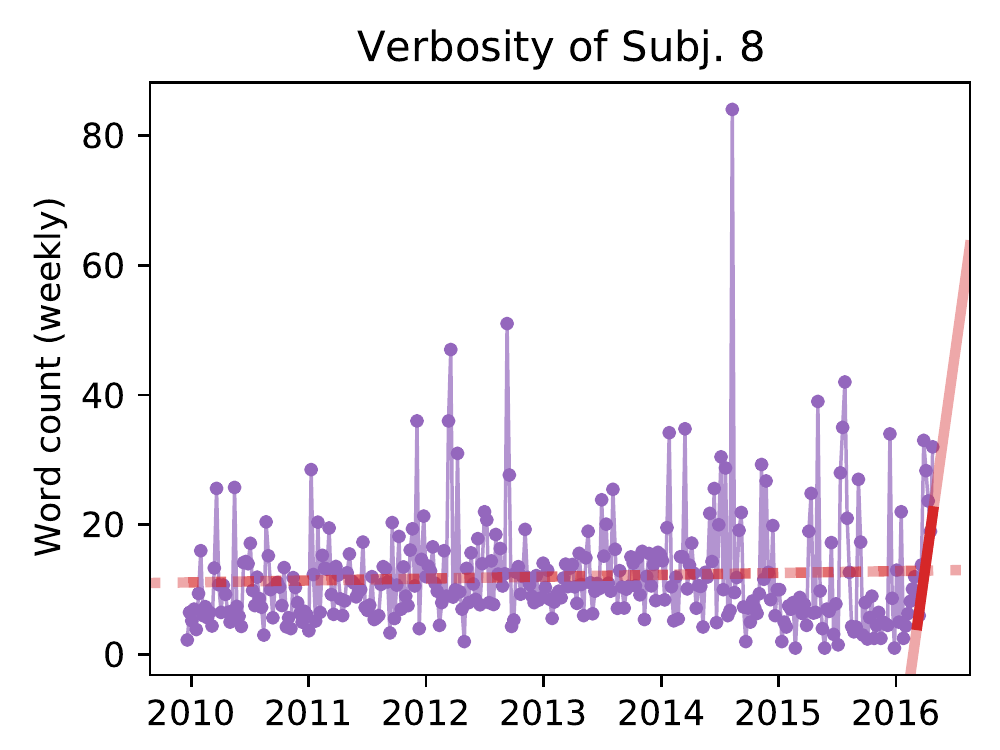}
    \includegraphics[width=.3\textwidth]{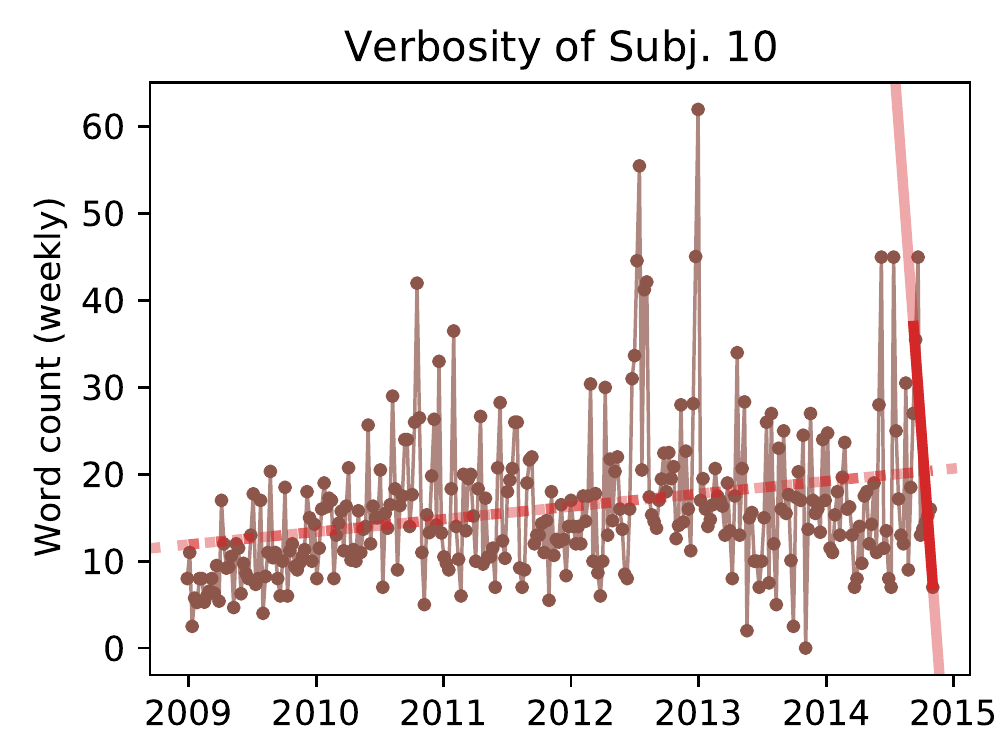}
    \caption{
        \textbf{Subject verbosity measured by word count}.
        Values are shown as weekly average to improve readability.
        Dashed red line shows the trend over the entire range of subject's posts.
        Solid red line is the trend over the last two months of data with darker color denoting the period length.
    }
    \label{fig:wordcount_regressions}
\end{figure}

Assuming some type of stressor prior to SUDEP, which in turn could manifest as a change in the subject's digital verbosity, first we characterize the number of words per subject \textit{Facebook} post (word count) with a simple negative binomial regression.
The binomial regression tests whether there was a significant difference in the amount of words per post when comparing posts written in two different epochs of the subject's digital behavior.
More specifically, we compare the average number of words per post in the two months (56 days) preceding the subject's SUDEP against the average number of words per post in the rest of the available timeline.
We choose the last two months as a conservative time range for a subject behavioral change that at the same time holds enough examples (posts) for a robust statistical analysis---as a 10 samples minimum is a frequently recommended heuristic for an accurate estimation of model parameters \cite{harrell2015regression}.
However we note that posting behavior varies between subjects and we do not know whether, or when, stressors proceedings SUDEP may appear for each subject.
We also tested different epochs, ranging from one to twelve weeks prior to SUDEP.
Results are consistent for subjects with sufficient data in the last period being considered, and are shown in Fig. \ref{fig:si:wordcount-sigtests-over-time}.
From our six analysed subjects, subjects 1, 2, 6, and 10 had significantly higher word count in the two months preceding their SUDEP.
Subject 8 also had a higher word count in the last two months, albeit not significant at $p<0.05$.
Conversely, subject 3 had a significantly lower word count in the last two months.
Results are shown in Table \ref{tab:wordcount-sigtests} and Figure \ref{fig:wordcount_regressions} shows the average word count for each subject timeline.
Two regressions are fitted to the data highlighting the slope of the increase (or decrease) in subject verbosity: one considering the complete subject timeline (dotted line) and one only considering the last two months of posts (solid line).

\begin{table}[ht!]
    \footnotesize
    \centering
    \begin{tabular}{l|r|r|r|r|r}
    \toprule
$subject$ & $n_{\textrm{early}}$ & $n_{\textrm{last}}$ & $\mu_{\textrm{early}}$ & $\mu_{\textrm{last}}$ & $time_{p}$\\
\midrule
2 & 2,162 & 109 & 12.431 & \textbf{34.413} & \textbf{1.197e-32} \\
1 & 1,547 & 54 & 9.592 & \textbf{17.889} & \textbf{4.146e-06} \\
8 & 2,185 & 16 & 12.070 & \textbf{18.375} & 0.081 \\
6 & 7,17 & 23 & 5.252 & \textbf{7.304} & \textbf{0.021} \\
\rowcolor{gray}
10 & 1147 & 7 & 13.983 & \textbf{23.571} & \textbf{0.048} \\
\rowcolor{gray}
3 & 834 & 10 & \textbf{11.125} & 4.100 & \textbf{0.001} \\
    \bottomrule
    \end{tabular}
    \caption{
        Significance tests for differences in word counts in posts during the last two months preceding SUDEP compared to other posts.
        The mean word count for the posts written during the last two months ($\mu_{\textrm{last}}$ with $n_{\textrm{last}}$ samples) are compared to the mean word count of all other posts written by the subject before this period ($\mu_{\textrm{early}}$ with $n_{\textrm{early}}$ samples). Significance is estimated from a negative binomial regression, with $p<0.05$ highlighted in bold.
        Subjects are ordered according to the rank-product of the number of samples during the last month and the number prior to the last month.
    }
    \label{tab:wordcount-sigtests}
\end{table}

Since digital behavioral changes may be reflected not only in post length but also in how frequent posts are made, next we use a zero-inflated negative binomial regression to examine whether the observed verbosity (word count) and frequency of posting prior to SUDEP was significantly different from subject's previous epochs.
A zero-inflated negative binomial regression is an extension of the binomial regression where there is an assumption that a different process governs the likelihood that a subject makes no posts in a day (zero word count), which is then modeled by a logistic regression.
Results are consisted and are presented in Table \ref{tab:si:wordcount-znbsigtests}; different epochs considered are shown in Figure \ref{fig:si:wordcount-znb-sigtests-over-time}.
In general we see that both subjects 1 and 2 were more likely to post in the two months preceding SUDEP, as well as writing longer posts.
Perhaps due to increased model complexity, changes in subject 6's posting behavior are less significant, being less likely to post in weeks preceding SUDEP with little difference in the number of words written per day.
Subject 8 and 10 were significantly less likely to post in the final weeks before SUDEP, with a non-significant increase in words per day when they did.
Lastly, subject 3 did not have a significant change in the number of days with a post, but did write significantly fewer words.

Having analyzed subject verbosity, we now turn to the sentiment of the text they wrote.
We remember each sentiment dimension is calculated by averaging per-word sentiment scores calculated for ANEW, LIWC, and VADER, three independent sentiment tools.
In the following Figures \ref{fig:valence-regressions}-\ref{fig:function-word-regressions}, line plots denote the average of a specific sentiment dimension measure over all posts each week.
Some particular sentiment trends can be observed in these figures.
For instance, four of the six subject show an over time increase in happiness sentiment, as measured by ANEW's valence dimension (see dotted lines in Fig. \ref{fig:valence-regressions}).
Only two subjects, 3 and 10, show a decrease in happiness in the last two months (solid line).
Importantly, Subject 3 has an overall happiness increase but the a sharp sentiment shift in the last two months, reflected by her described feelings of loneliness of being a college freshmen.
On the other hand, subject 6 has an over time happiness decrease, but a sharp happiness increase in the last two months, reflecting a sentiment shift due to her European travels.
Overall, despite some subjects having reversed valence sentiment, when their complete timeline sentiment is compared to the sentiment in the last two months of posting, they all have something in common: a significant sentiment shift, as measured by the difference in slope of the two regressions.

To show this phenomena is not simply an effect of the sentiment tool of choice, Figures \ref{fig:neu-regressions} \& \ref{fig:function-word-regressions} show subject use of emotion-neutral words and functional words, measured by VADER and LIWC, respectively.
Functional words includes a broad category of words such as pronouns (`him', `she'), articles (`the', `a'), conjunctions (`and', `but'), interjections (`oh', `ah'), pro-sentences (`yes', `no', `okay'), and others.
We observe an over time increase in the average number of such words used per post for 5 of the 6 subjects (see Fig. \ref{fig:function-word-regressions}). In addition, for 4 subjects the amount of functional words used increases substantially in the last two months of posting.
In regards to emotion-neutral words, five of the six subjects show an increase use of emotion-neutral words---a sentiment dimension that other tools, such as ANEW, ignores (see Fig. \ref{fig:neu-regressions}).
However, similarly to subject verbosity, all subjects have a drastic shift in the analyzed sentiment categories when their complete timeline is compared to the last two months, again as measured by the regression slope (see red lines in aforementioned plots).

\begin{figure}[ht!]
    \centering
    \includegraphics[width=.3\textwidth]{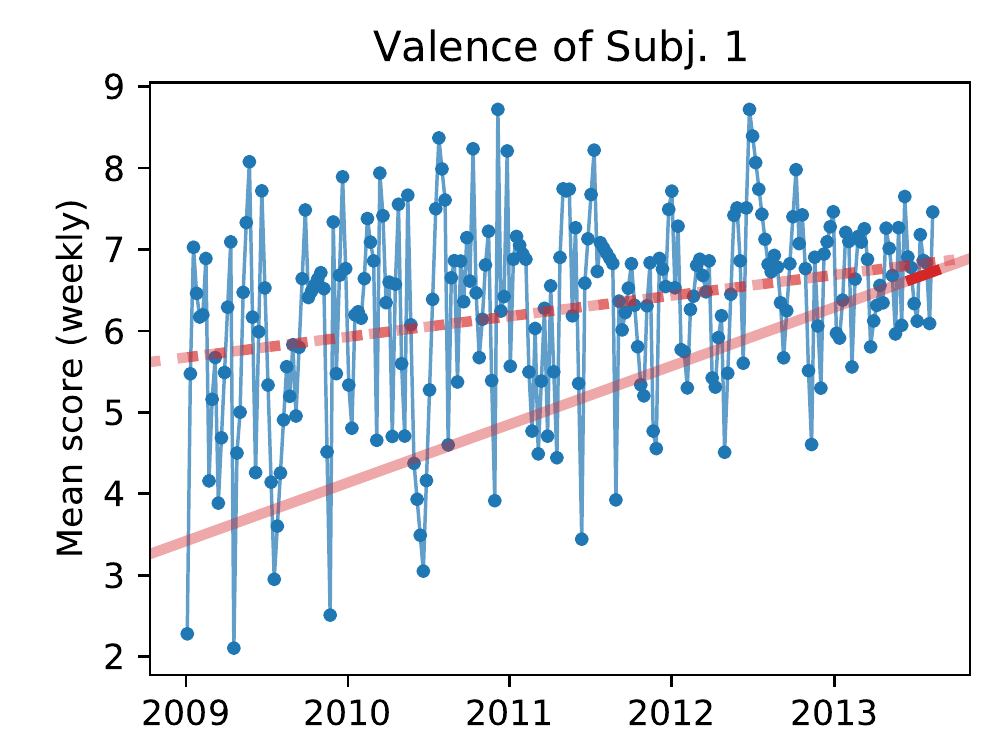}
    \includegraphics[width=.3\textwidth]{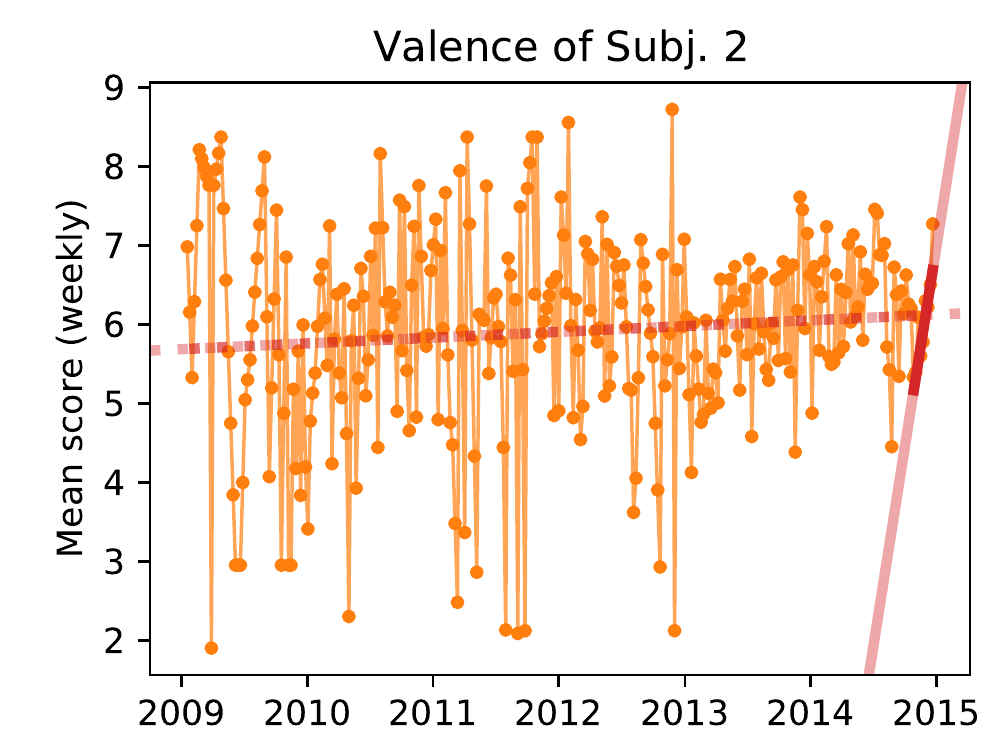}
    \includegraphics[width=.3\textwidth]{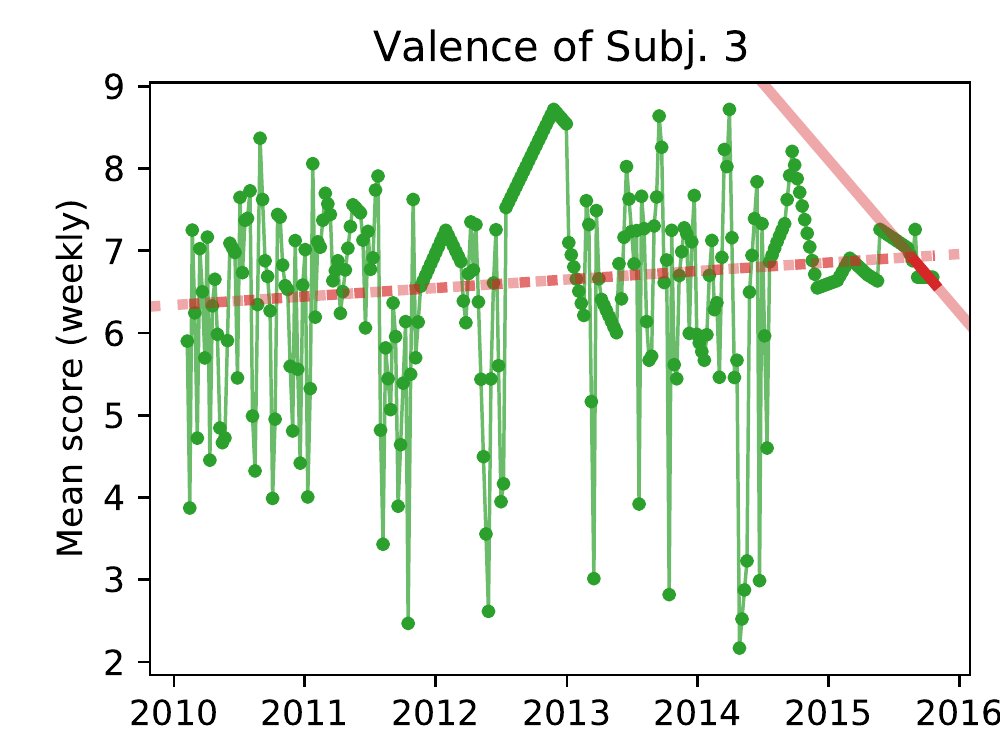}
    \includegraphics[width=.3\textwidth]{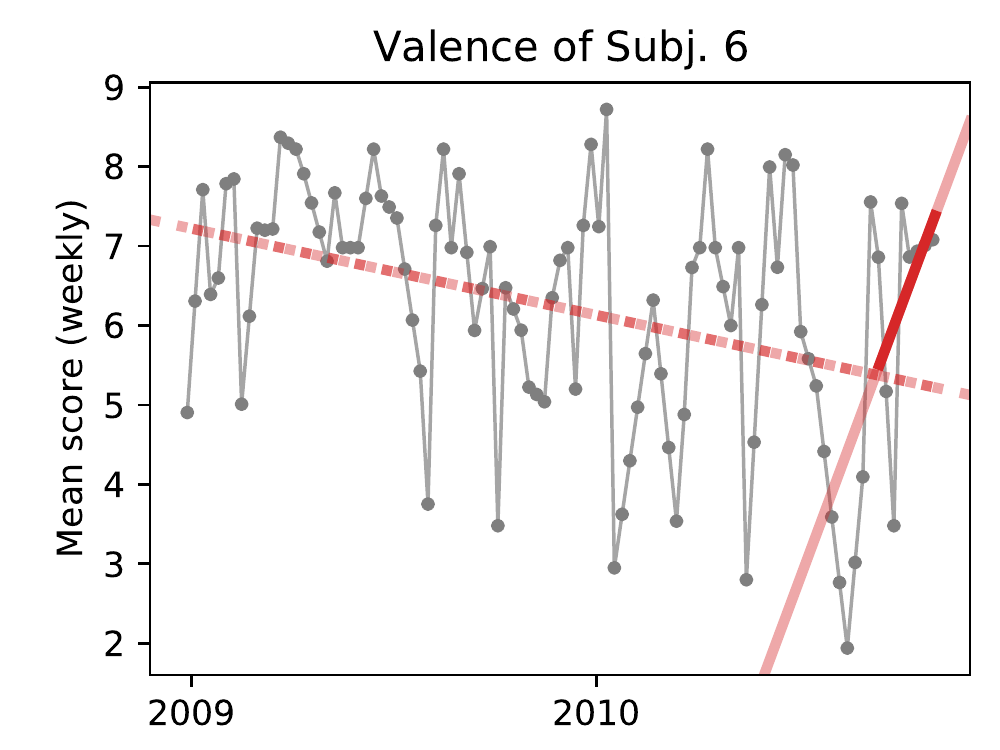}
    \includegraphics[width=.3\textwidth]{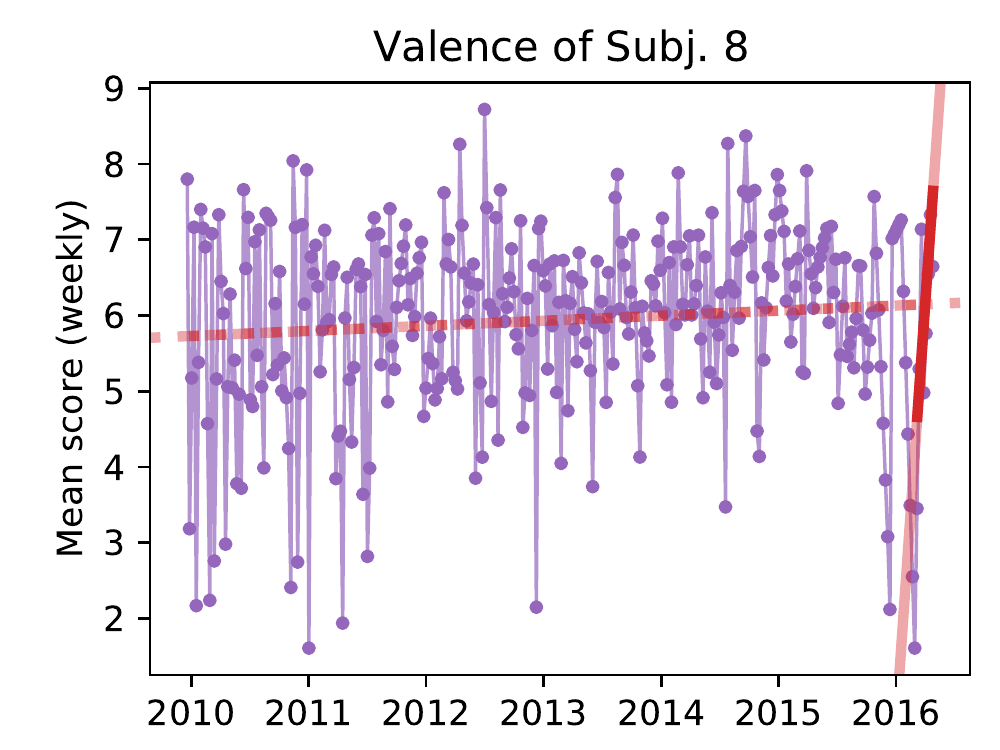}
    \includegraphics[width=.3\textwidth]{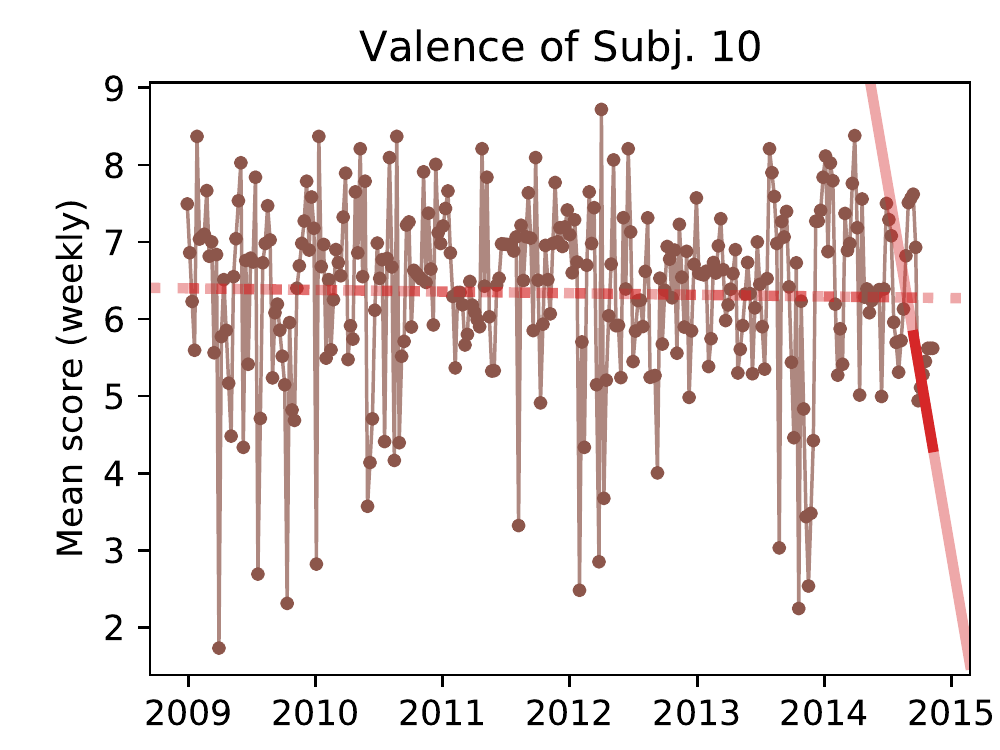}
    \caption{
        \textbf{Subject happiness measured by ANEW's Valence score}.
        Values are shown as weekly average to improve readability.
        Dashed red line shows the trend over the entire range of subject's posts.
        Solid red line is the trend over the last two months of data with darker color denoting the period length.
    }
    \label{fig:valence-regressions}
\end{figure}

\begin{figure}[ht!]
    \centering
    \includegraphics[width=.3\textwidth]{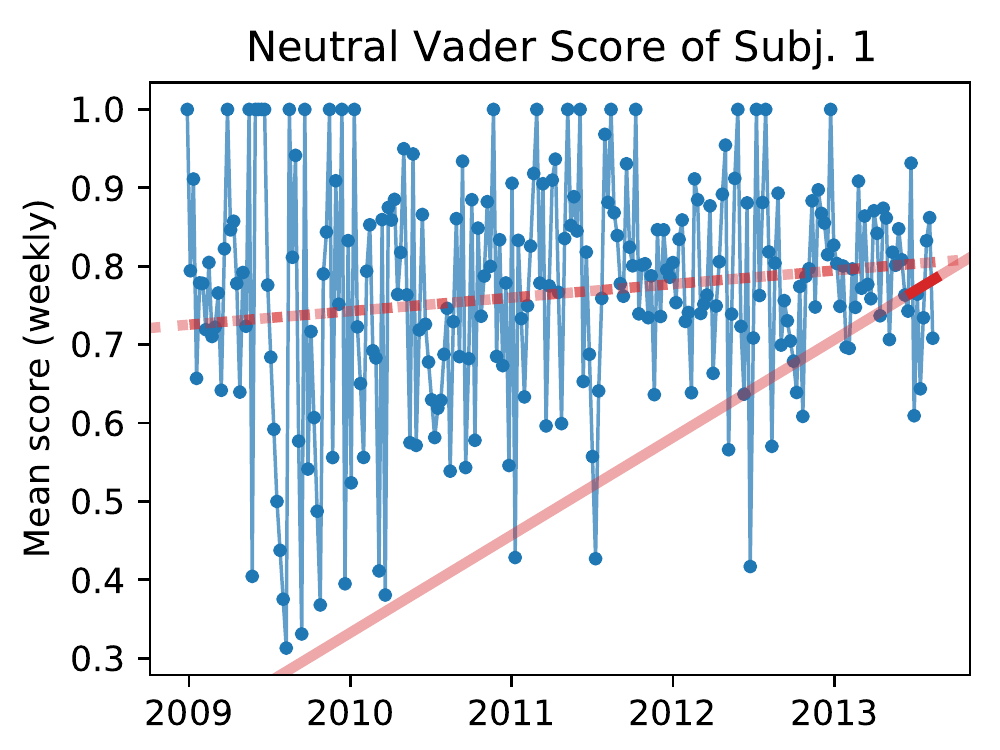}
    \includegraphics[width=.3\textwidth]{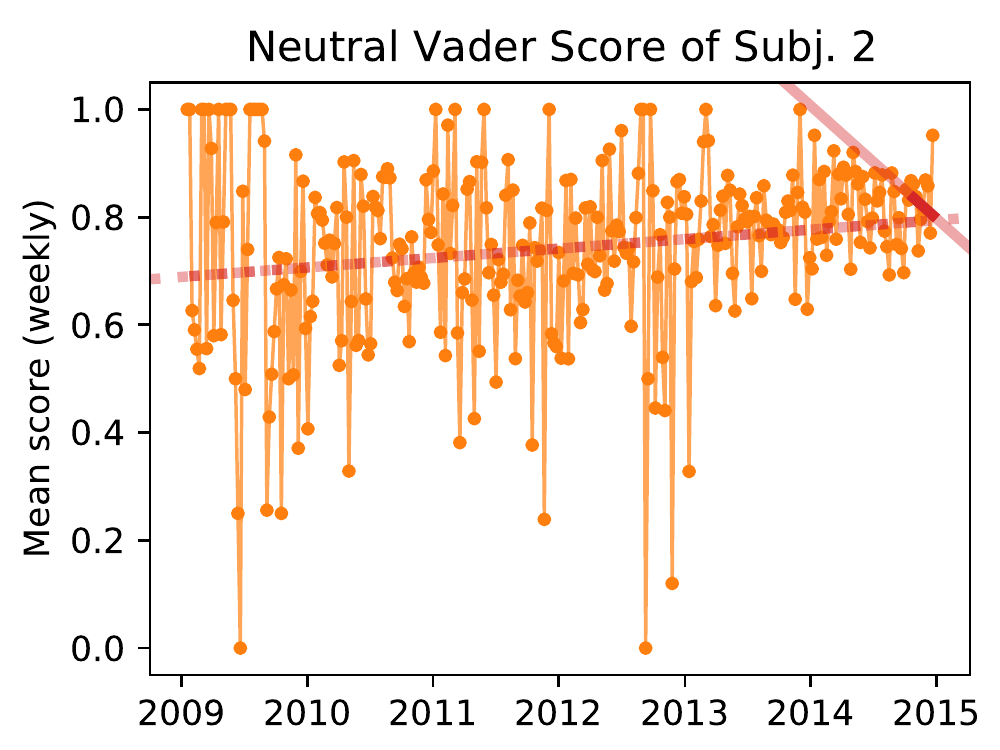}
    \includegraphics[width=.3\textwidth]{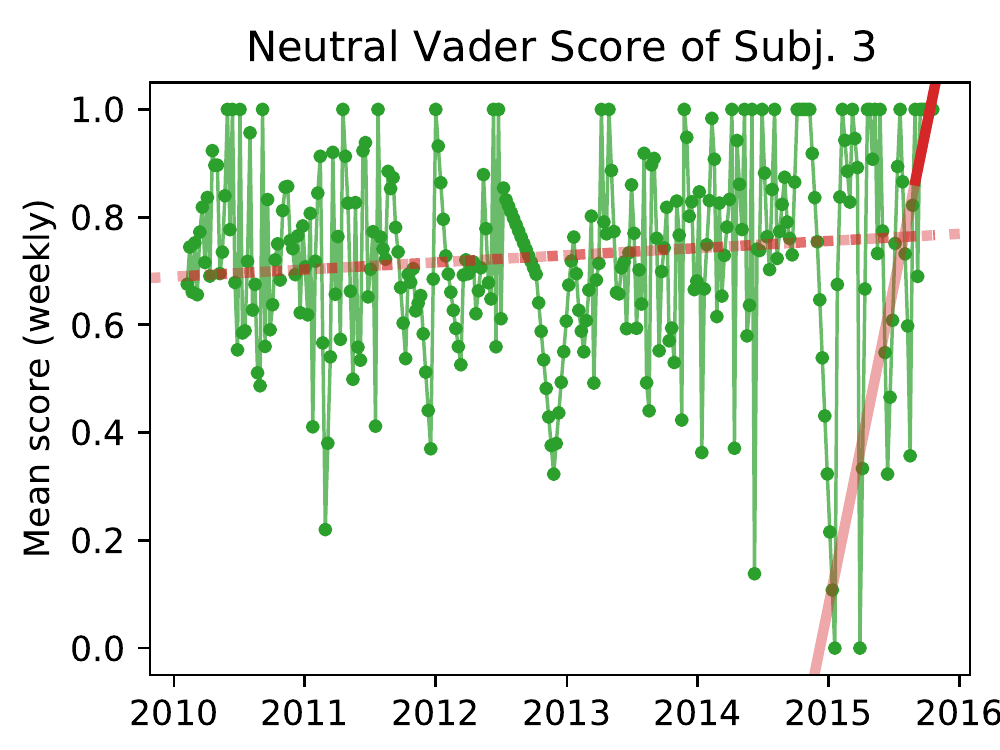}
    \includegraphics[width=.3\textwidth]{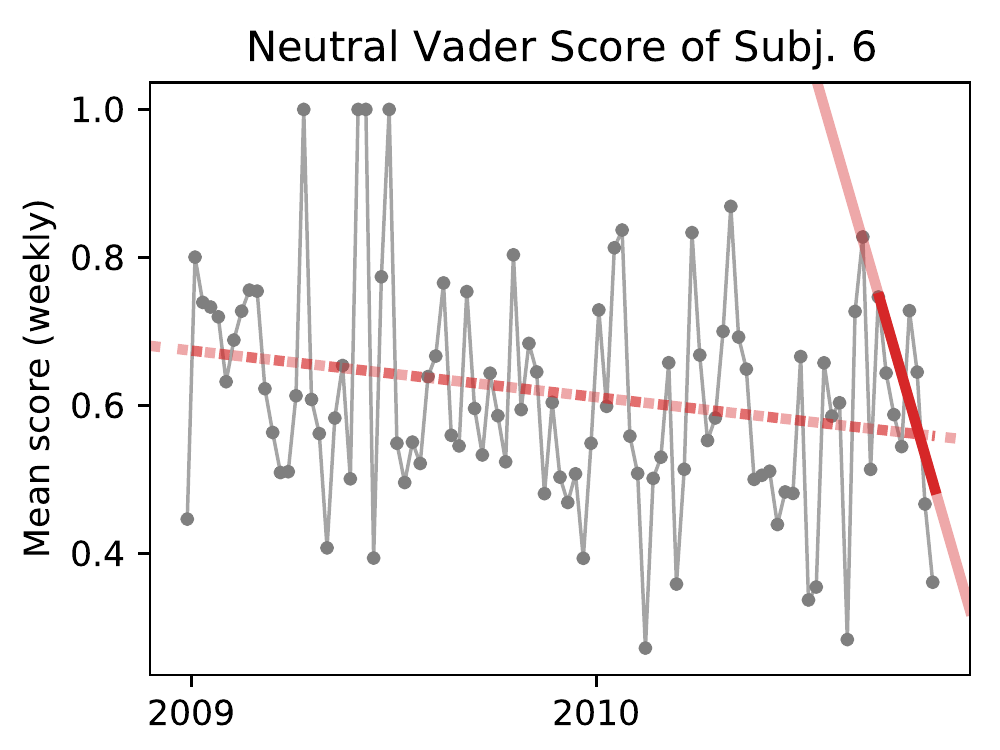}
    \includegraphics[width=.3\textwidth]{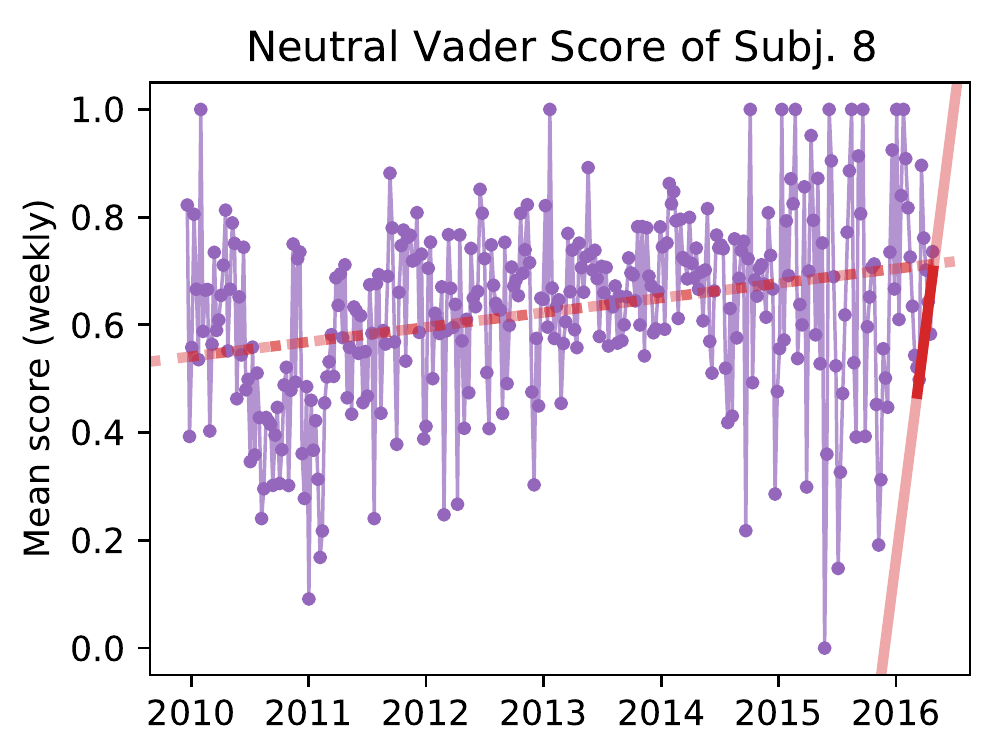}
    \includegraphics[width=.3\textwidth]{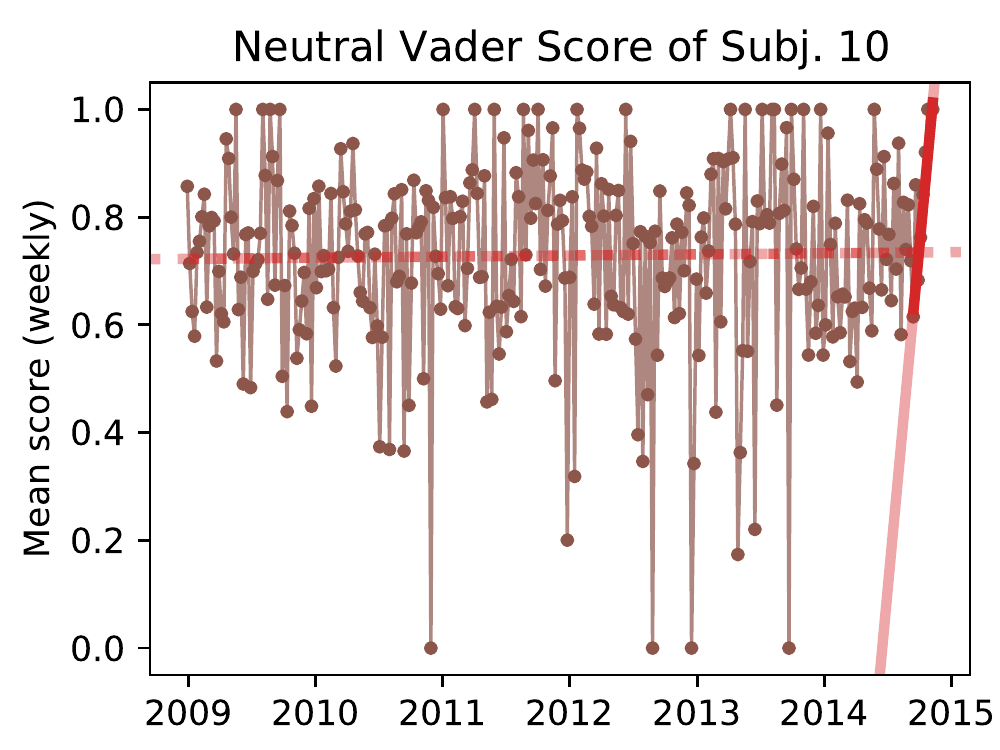}
    \caption{
        \textbf{Subject use of neutral words measured by VADER's Neutral score}.
        Values are shown as weekly average to improve readability.
        Dashed red line shows the trend over the entire range of subject's posts.
        Solid red line is the trend over the last two months of data with darker color denoting the period length.
    }
    \label{fig:neu-regressions}
\end{figure}

\begin{figure}[ht!]
    \centering
    \includegraphics[width=.3\textwidth]{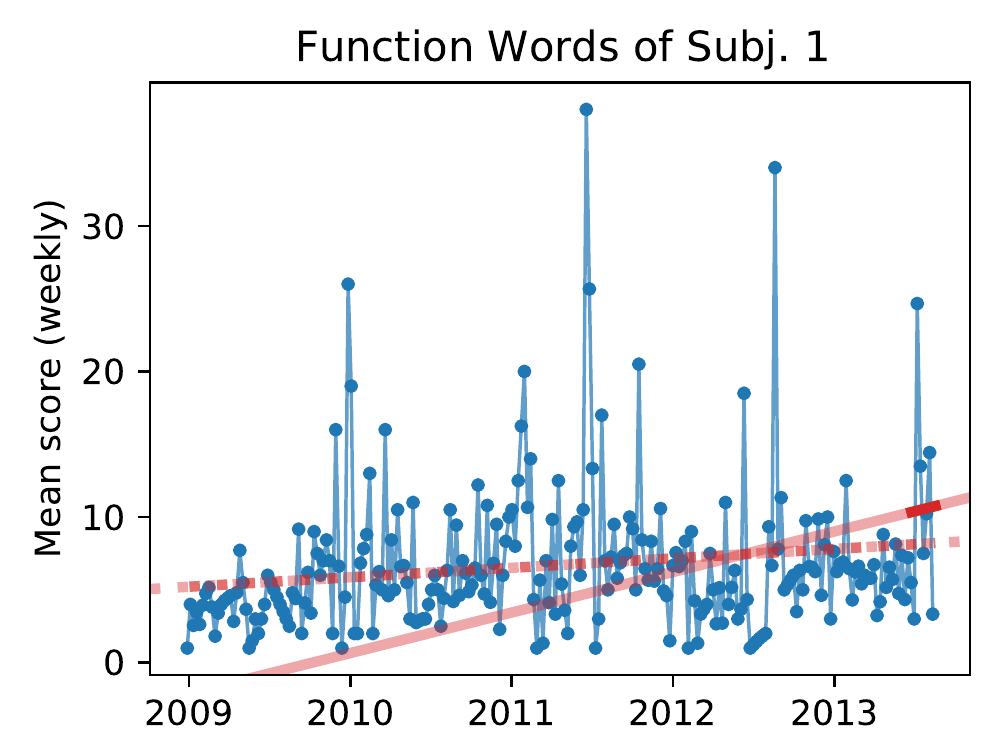}
    \includegraphics[width=.3\textwidth]{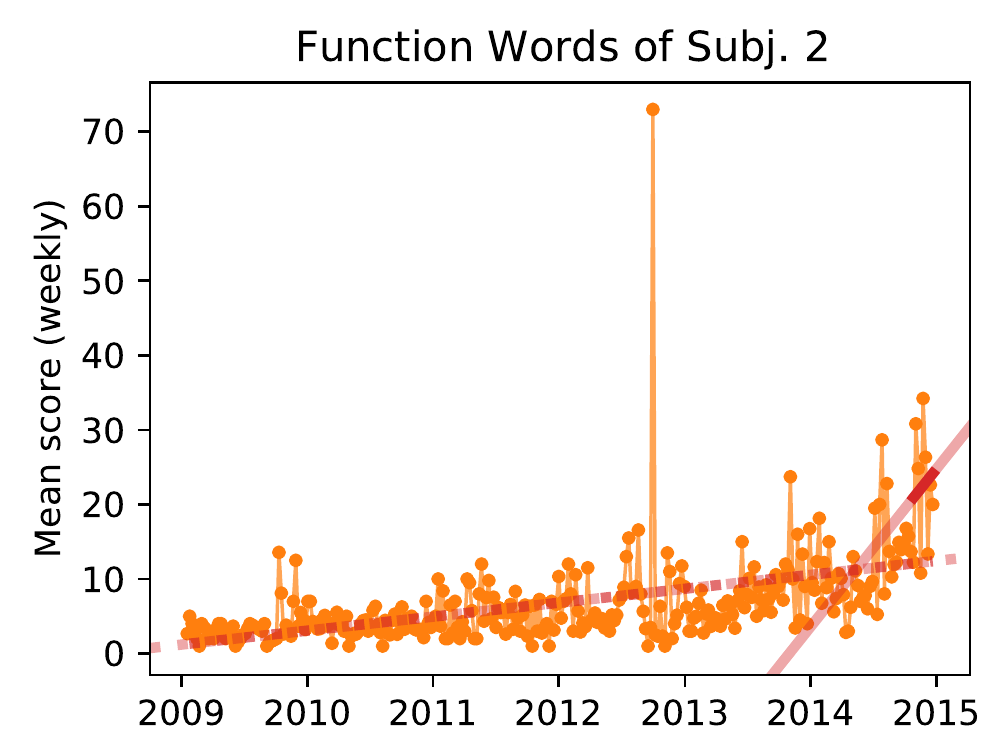}
    \includegraphics[width=.3\textwidth]{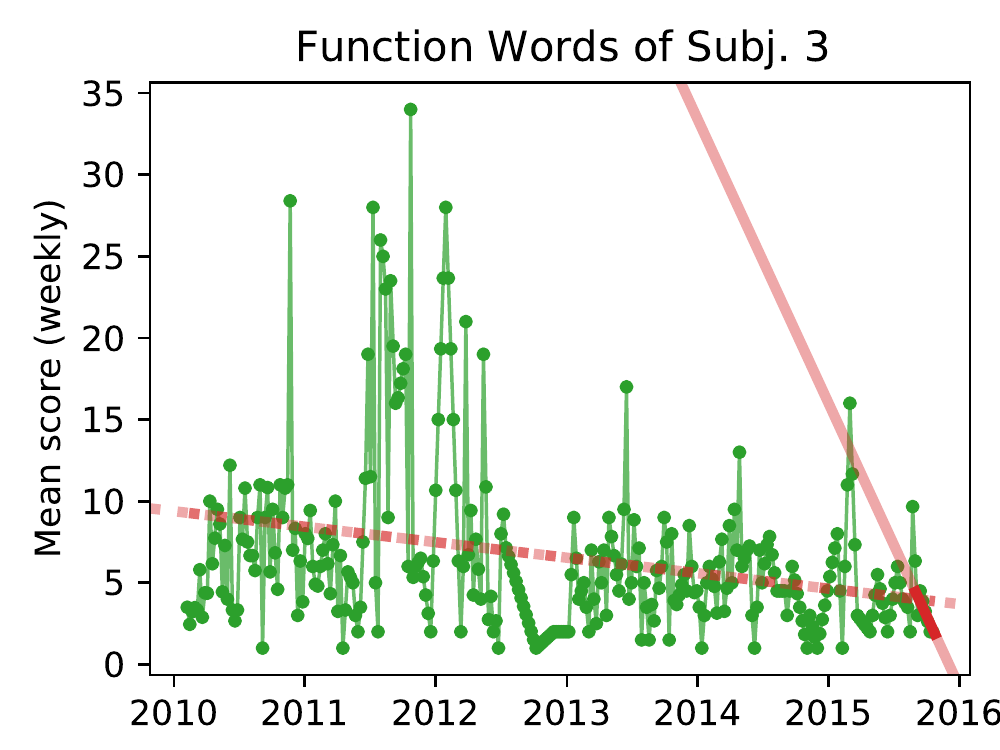}
    \includegraphics[width=.3\textwidth]{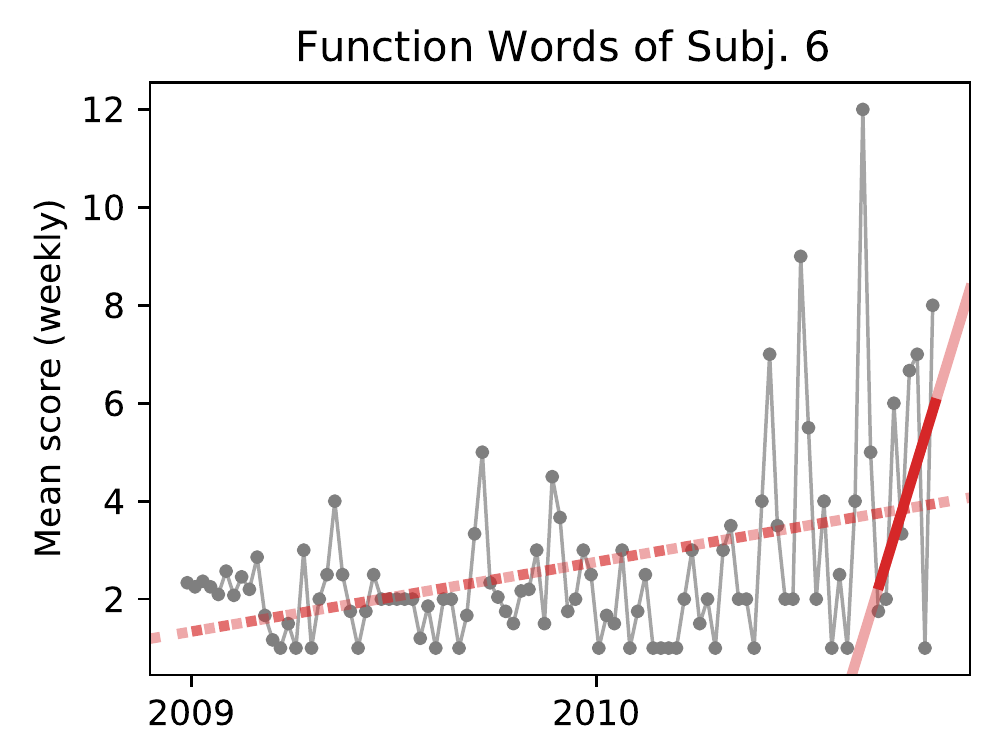}
    \includegraphics[width=.3\textwidth]{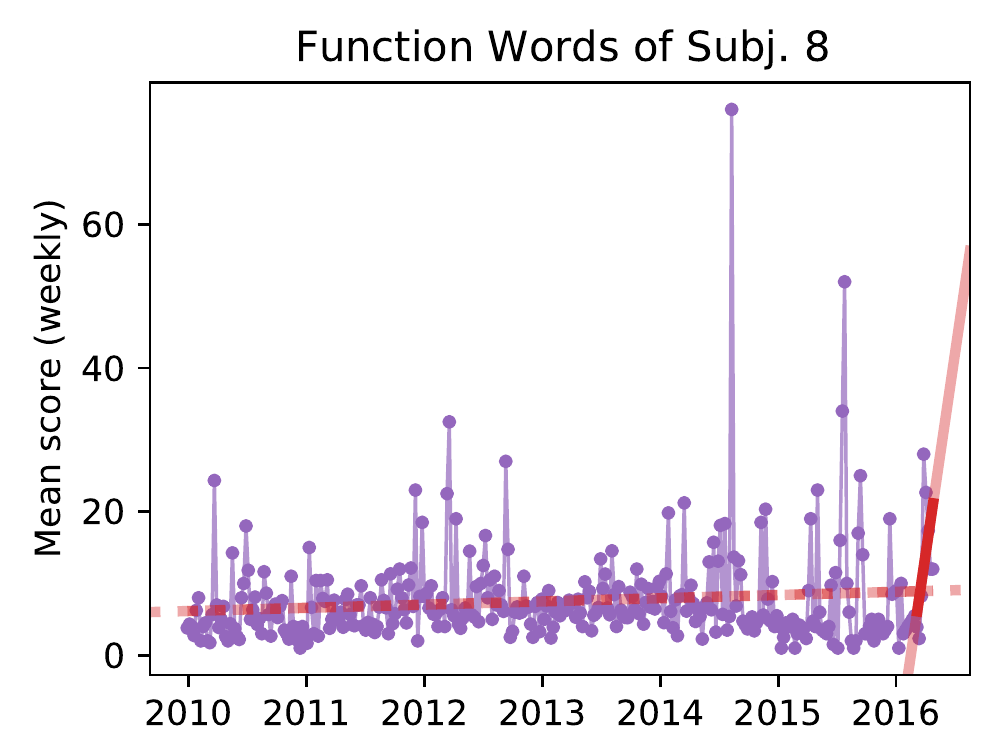}
    \includegraphics[width=.3\textwidth]{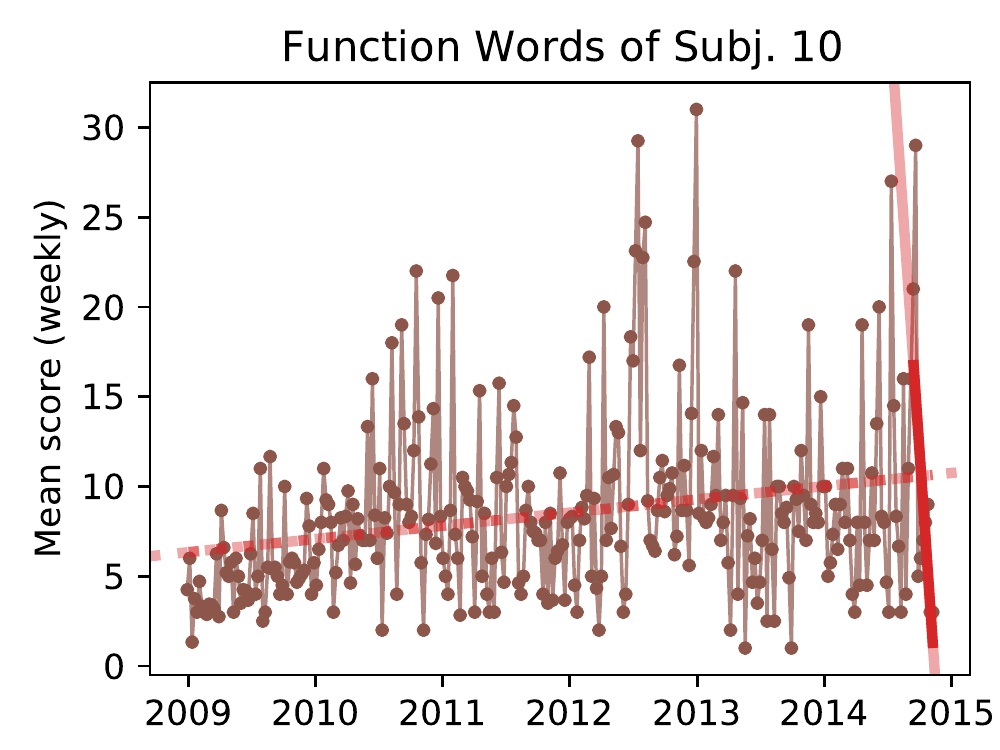}
    \caption{
        \textbf{Subject use of functional words measured by LIWC}.
        Values are shown as weekly average to improve readability.
        The dashed red line shows the trend over the entire range of a subject's posts, while the solid red line is the trend over the last two months of data.
    }
    \label{fig:function-word-regressions}
\end{figure}

\section{Discussion}
\label{ch:discussion}

First, we would like to emphasize that we cannot claim SUDEP causation, or the predictive accuracy of these tools applied to the social media posts of living individuals.
However, the noticeable increase in functional words and the overall verbosity preceding SUDEP for a number of subjects is particularly suggestive of some detectable changes in the digital behavior of subjects, and that may serve as early-warning signals correlating with SUDEP.
It is known that stress and major life events are likely to increase the risk of epilepsy \cite{McConnell:1995, McKee:2017}, and that in turn may increase the risk of SUDEP.
Several of our subjects had major life changes in the weeks preceding their death, from concussions, moving to another city, returning from an overseas trip, or feeling lonely as a new college freshmen.
In addition, the misuse of functional words has been associated with Aphasia, a language impairment attributed to the Wernicke's area, a brain area in the left (dominant) temporal-parietal region characterized by EEG abnormalities in epilepsy patients \cite{Chung:2007,StatPearls:2020:wernicke-aphasia,EFA:2021:language-problems}. Unlike impairment to Broca's area where patients speak slow, in hesitating ways, and phrases are devoid of functional words, impairment to the Wenicke's area cause patients to speak warmly and fluidly but using functional words with no content at all \cite{Chung:2007}.
We manually checked sentence construction in the last two months of posting for our subjects and found no trace of functional words misuse aside from their increased occurrence.
Nonetheless, if an increase in verbosity or changes in functional word use is indicative of stress or major life changes, the use of textual and sentiment tools may allow for a predictive, quantitative measure in larger studies, complementing current qualitative analyses.
But we do stress that the lack of appropriate sample size and a rigorous case-control in our current study hinders generalization of our findings at this point.
Nonetheless, our preliminary results serve to invite additional research into this problem, especially to encourage attention to social media and other digital behavior data, thus contributing to better prediction of warning signals of SUDEP.

One possible avenue to evaluate the  potential of sentiment analysis for predicting SUDEP is to employ statistical machine learning models using the text and sentiment analysis tools we described above.
We attempted to build such models to predict changes in the last day or week of posts in a subject's timeline---instead of the last two months of posts shown in regressions above.
However, we encountered two common machine-learning problems, especially in shorter window scenarios.
The first was over-fitting and the subsequent false positive prediction.
Since sentiment tools possess many sentiment variables (dimensions), it is easy to perfectly fit posts used in training the algorithm.
Yet, the resulting prediction/classification models do not generalize to predicting subject posts left out for testing.
Stricter model regularization and dimensionality reduction methods can help, but in the end, using shorter prediction windows results in a classification scenario with a very large class imbalance with very few positive instances (i.e., posts preceding SUDEP) which does not allow automatic machine learning classification.
This is because most posts occur when subjects are deemed healthy, and only very few instances can be safely set as being SUDEP related---those that happened right before death.
Given this problem of class imbalance, classifiers for automatic prediction are not possible with our current dataset. 

The second problem pertains to the labeling of posts as SUDEP-relevant. Assuming that only the last posts before SUDEP are relevant, may miss prior days and posts (positive instances) that may have been  close calls for SUDEP.  Without the proper labeling of these instances, our algorithms are potentially missing several learning opportunities.
The two-month window prior to SUDEP we used in the regression analysis is reasonable for the observed cohort, allowing a reasonable amount of positive posts for most subjects (see Table \ref{tab:si:wordcount-znbsigtests}). But the regression serves as an observation tool more than an automatic predictor.
Indeed, at the current stage, social media analysis can only enhance and provide a different perspective to other health data, such as electronic health records, personal diaries, epilepsy warning devices, service animals, etc.
A more systemic and complete picture of SUDEP may emerge by combining these seemingly heterogenous data sources.

Going forward, our goal is to combine clinical (e.g., physician notes, laboratory exams, genetic profiling, questionnaire responses, electronic health records) with non-clinical digital behavioral data (e.g. electronic diaries, discussion boards, email exchange, phone usage patterns, social media posting and consumption) into research design. This is planned via recruitment of epilepsy patients who consent the to the collection of their digital behavioral data, such as social media IDs \cite{Correia:2020}. Our own work with focus groups of epilepsy patients and their caretakers have demonstrated willingness to donate digital behavioral data for studies. Indeed, as shown in the work we report here, this can be even done postmortem to avoid an observer bias---patients changing their behavior by knowing they are being observed. With enough subjects to account for the increase in variables, the next step is to validate the predictive power of social media signals in case-control experiments. We intend to focus on specific questions such as: why are subjects writing or using certain words more often prior to their death? Can this be statistically correlated with an increased risk of SUDEP? Can we pinpoint a behavioral phase shift to inform self- and caretaker-management as an early warning? The preliminary results we now report demonstrate the feasibility of extracting such signals.
As we recruit additional subjects in planned larger studies, it will be possible to answer these questions more quantitatively and conclusively.

To compile additional digital behavioral data sources, our team is currently developing myAura \cite{myAURA}, a personalized web service for epilepsy management. MyAura will include self-reported patient diaries, such as seizure tracking, food and water intake, medication adherence, physician encounters, among others. One of its goals is to test a variety of clinical and non-clinical temporal variables that may be proven useful in epilepsy management.
The use of patient donated social media timelines, as we have shown here, can prove to be the next frontier in informing our understanding of SUDEP and other epilepsy outcomes.
MyAura will include the option for users to donate their social media timelines, thus allowing the recruitment of larger patient cohorts.
Findings from analysis of the data of larger cohorts is likely to inform self-management recommendations for PWE, including allowing for SUDEP-predicting behaviors to be identified. For instance, patients with epilepsy could be monitored for an increased risk for SUDEP. In addition, our text and sentiment analysis could be used to inform individualized self-management interventions based on patient's posts and behaviors.
At the same time behavioral results can help direct physiologic studies, as cellular-level or biomarker changes can, for example, ultimately be correlated with behavioral experiences (e.g. cortisol and physiologic or psychological stress).

As a small pilot, our study has demonstrated the feasibility of mining social media data for SUDEP (and other epilepsy-related) research, as well as very preliminary findings regarding increased social media activity preceding SUDEP.
While the sample size of this study is too small to render generalizations in terms of SUDEP prediction, our work here demonstrates the feasibility of a novel way of investigating epilepsy-related phenomena, including SUDEP.
This work also demonstrates the value in the interdisciplinary collaboration between clinical/behavioral epilepsy researchers and informatics/complex systems scientists.

\section*{Acknowledgments}
\label{ch:ack}
\footnotesize

This research was funded by the National Institutes of Health, National Library of Medicine Program, grant 1R01LM012832-01, as well as  by the Indiana Clinical Translational Sciences Institute, grant NIH/NCRR UL1TR001108.
RBC was also a CAPES Foundation fellow, grant 18668127.
The funders had no role in study design, data collection and analysis, decision to publish, or preparation of this manuscript.

%
% References
%
\bibliographystyle{elsarticle-num}
\bibliography{FBEpilepsy}

\pagebreak
\clearpage

\appendix
\renewcommand\thefigure{S\arabic{figure}}
\renewcommand\thetable{S\arabic{table}}
\setcounter{figure}{0}
\setcounter{table}{0}

%
% Supplemental Materials
%
\section*{Supplementary Materials}

\begin{figure}
    \centering
    \includegraphics[width=.45\textwidth]{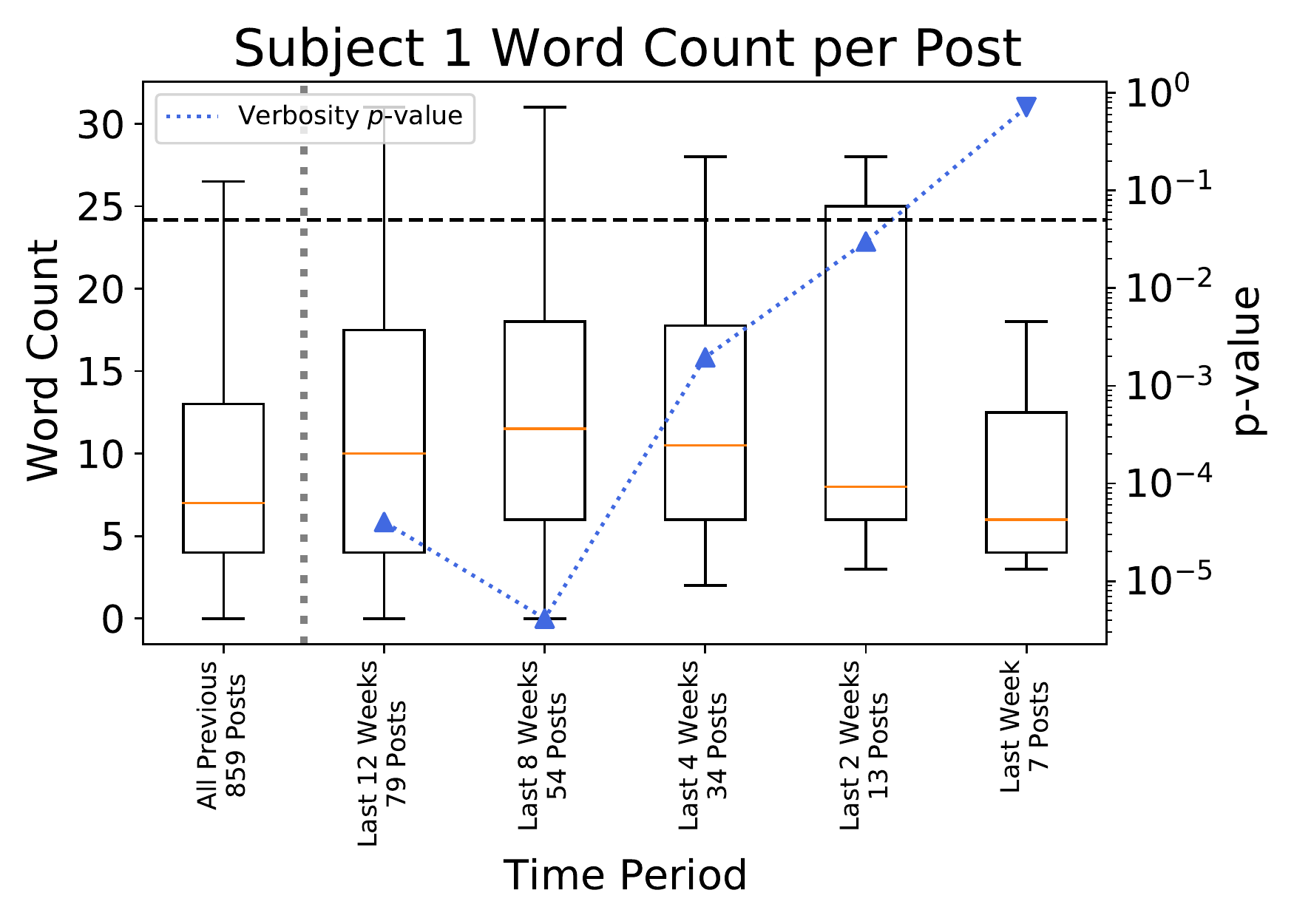}
    \includegraphics[width=.45\textwidth]{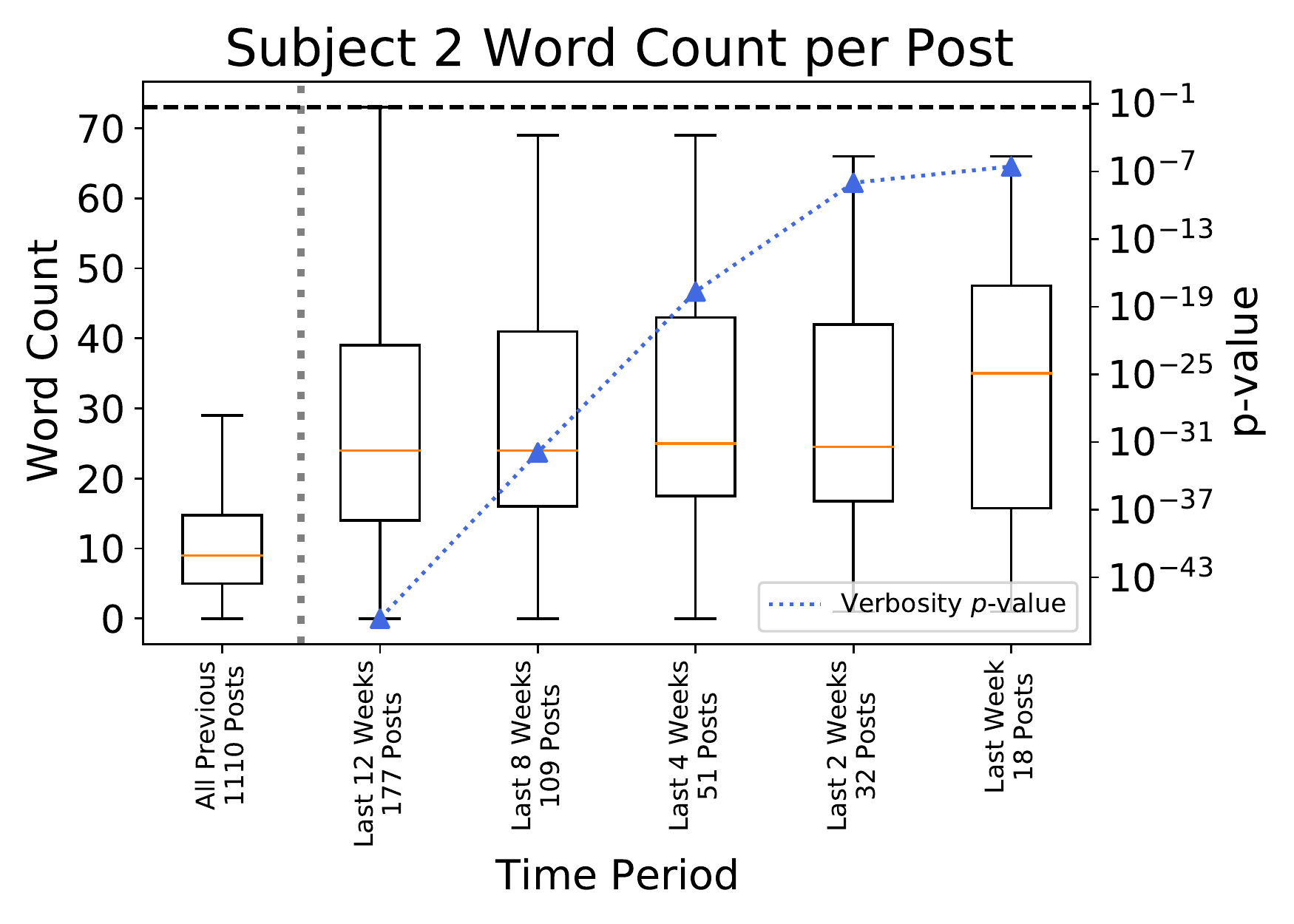}
    \includegraphics[width=.45\textwidth]{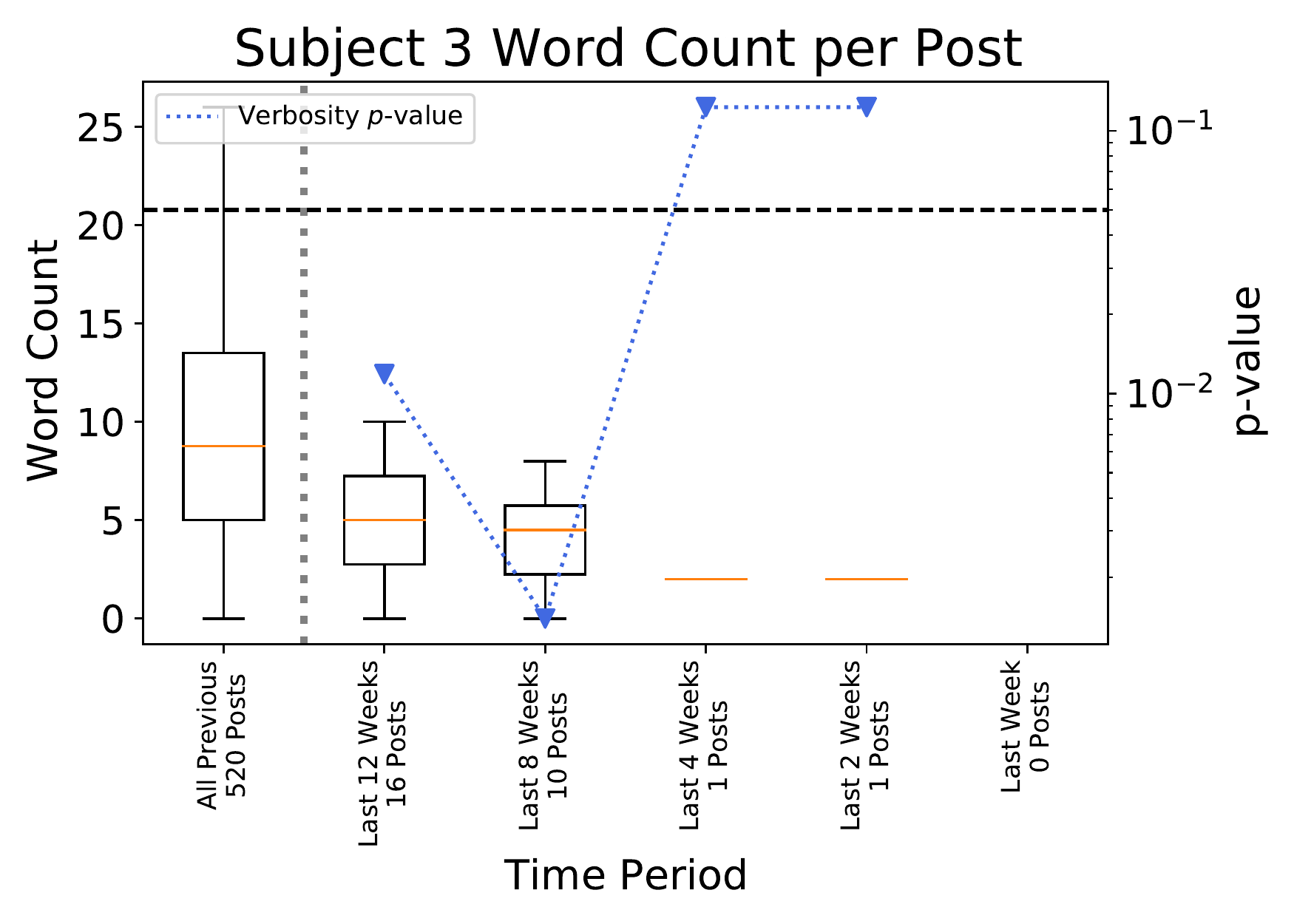}
    \includegraphics[width=.45\textwidth]{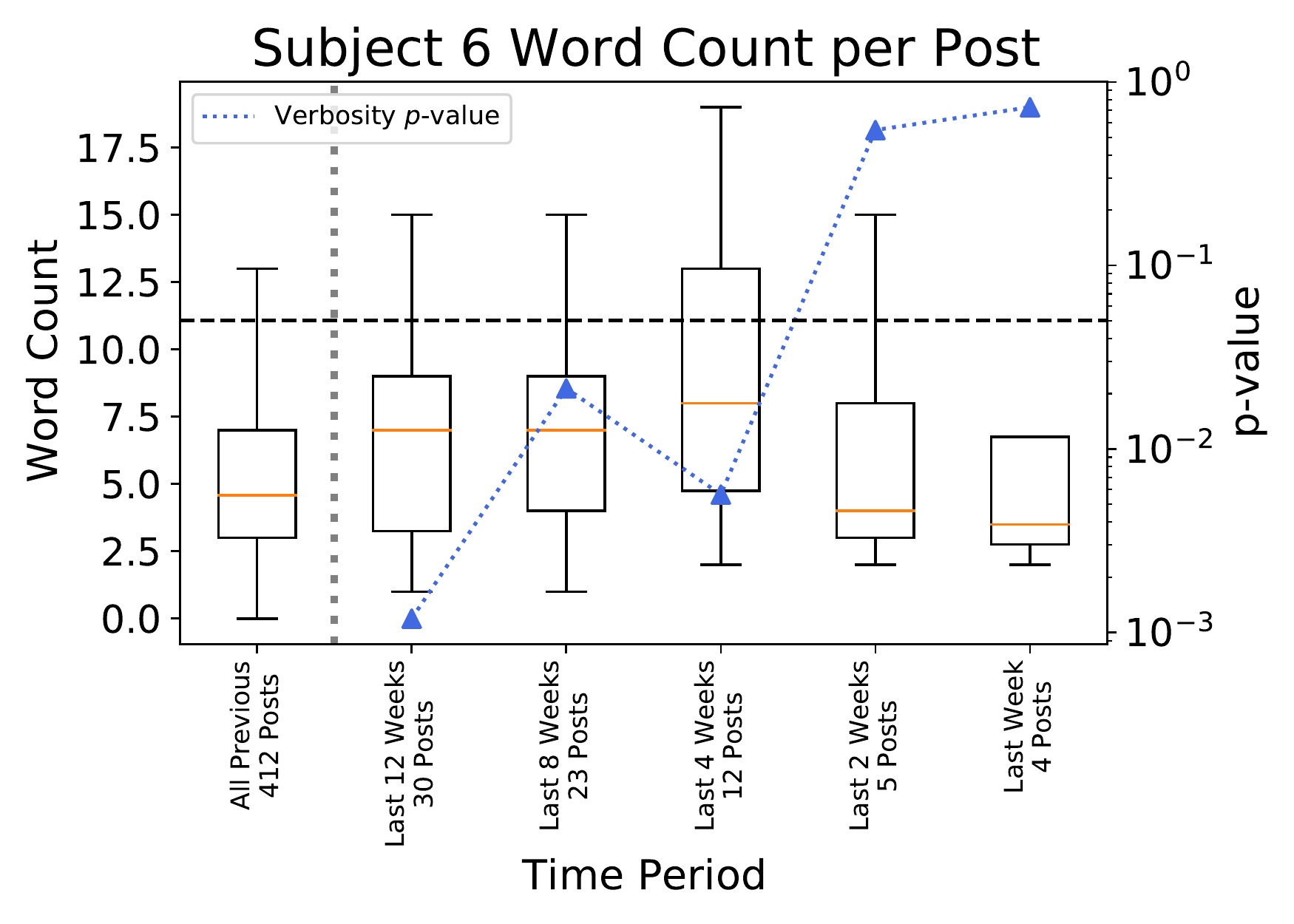}
    \includegraphics[width=.45\textwidth]{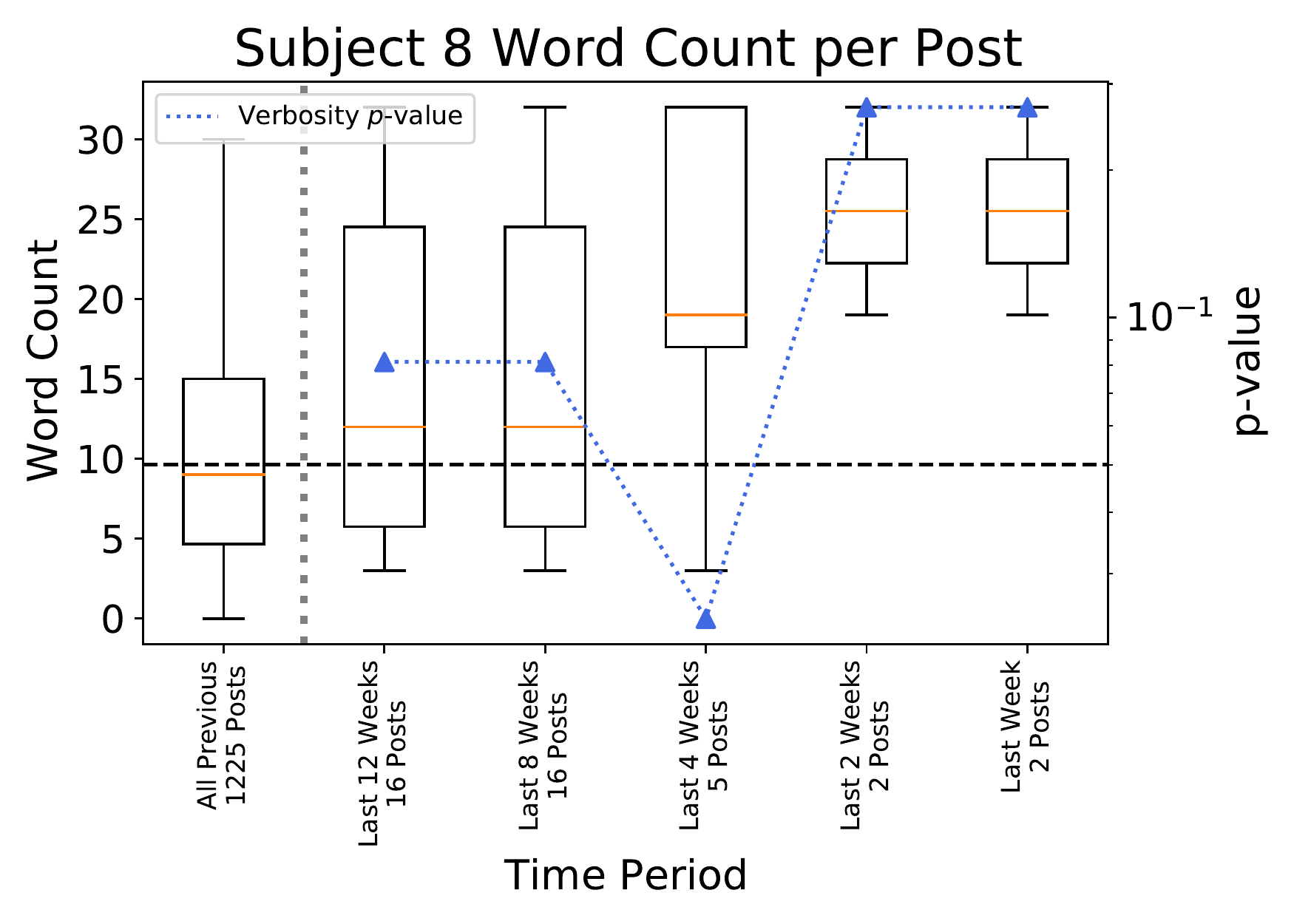}
    \includegraphics[width=.45\textwidth]{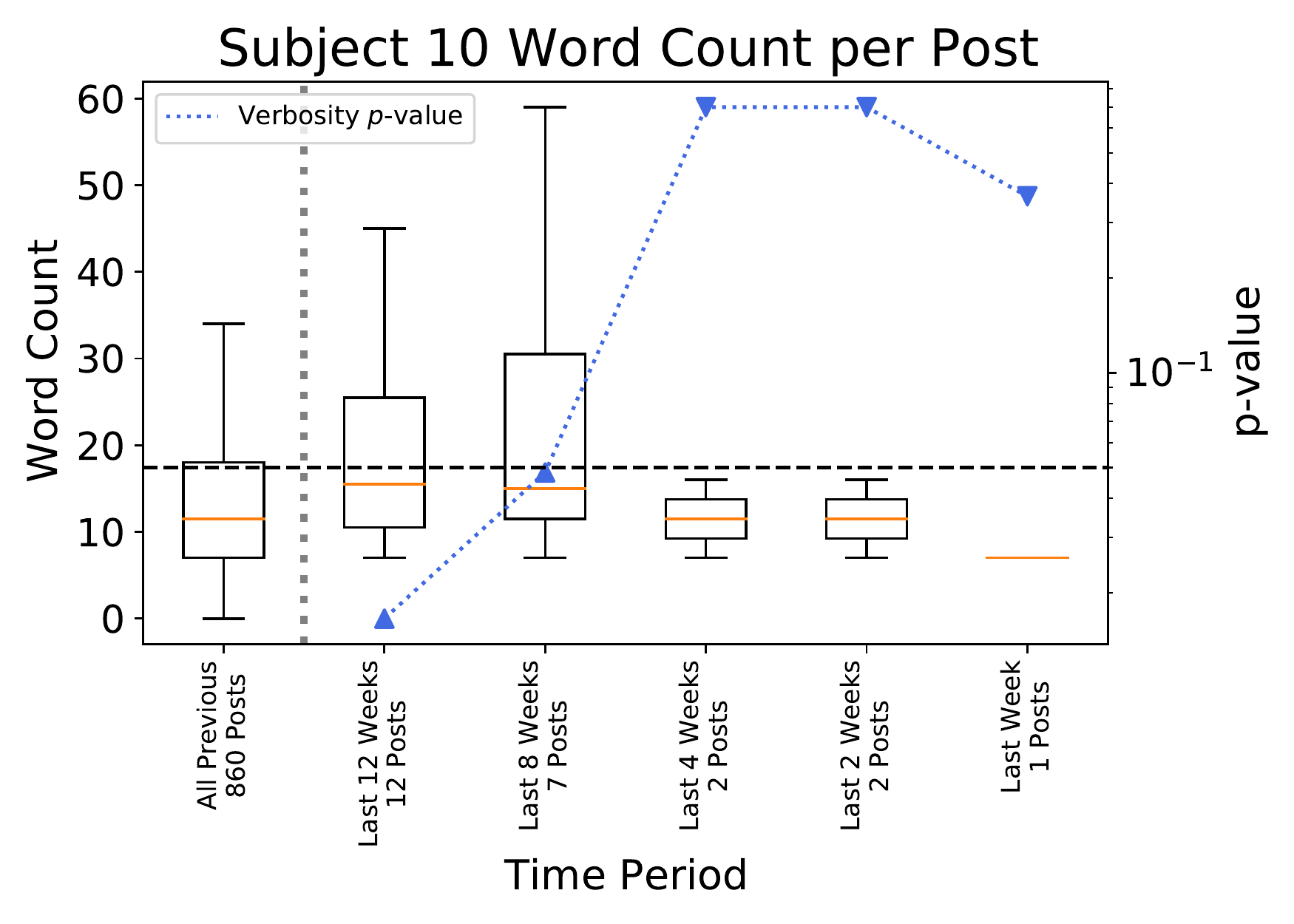}
    \caption{
        \textbf{Subject verbosity per post over different epochs}.
        Difference between word count per post in the period immediately preceding SUDEP compared to word count per post during earlier posting periods.
        Different selections of the time window for the last posting period are displayed on the x-axis.
        The box plot on the far left represents all posts before the 12 weeks preceding SUDEP.
        The blue line represents the p-value of the time coefficient for the negative binomial regression.
        The direction of the arrow represents the sign of the coefficient, up indicates an increase in wordcount during the period preceding SUDEP and down indicates a decrease.
        The horizontal black line represents p=0.05
    }
    \label{fig:si:wordcount-sigtests-over-time}
\end{figure}

\begin{figure}[!h]
    \centering
    \includegraphics[width=.45\textwidth]{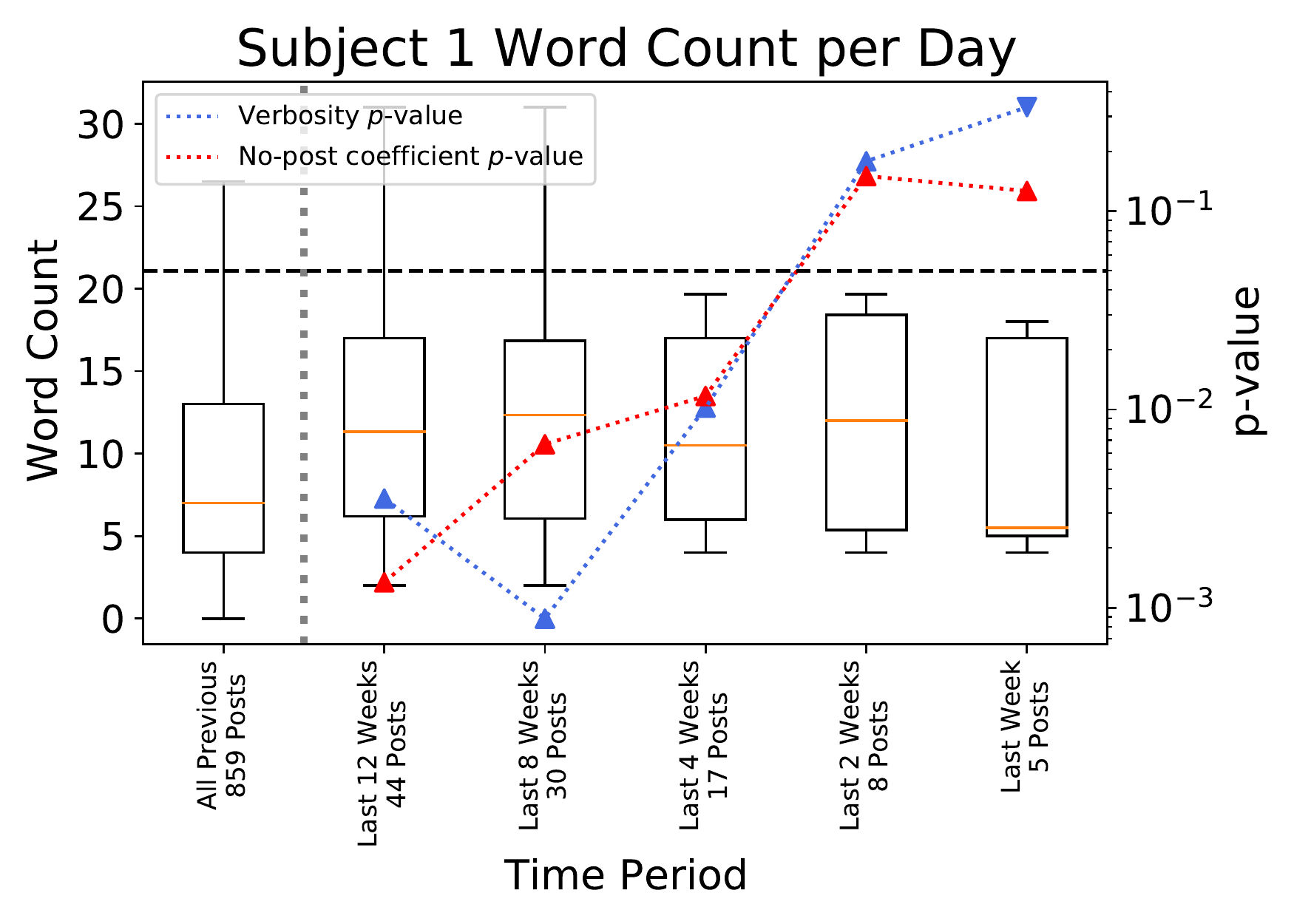}
    \includegraphics[width=.45\textwidth]{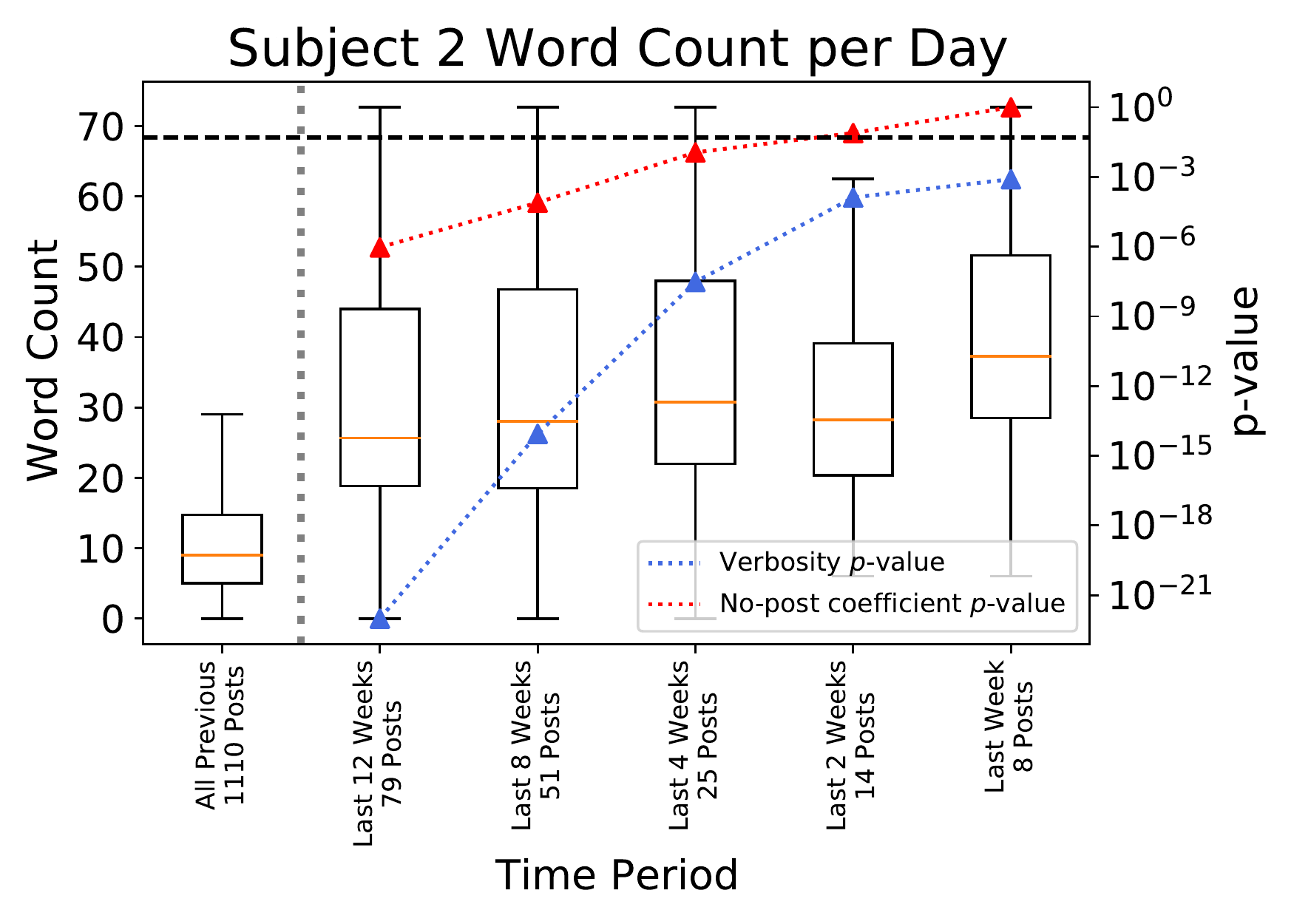}
    \includegraphics[width=.45\textwidth]{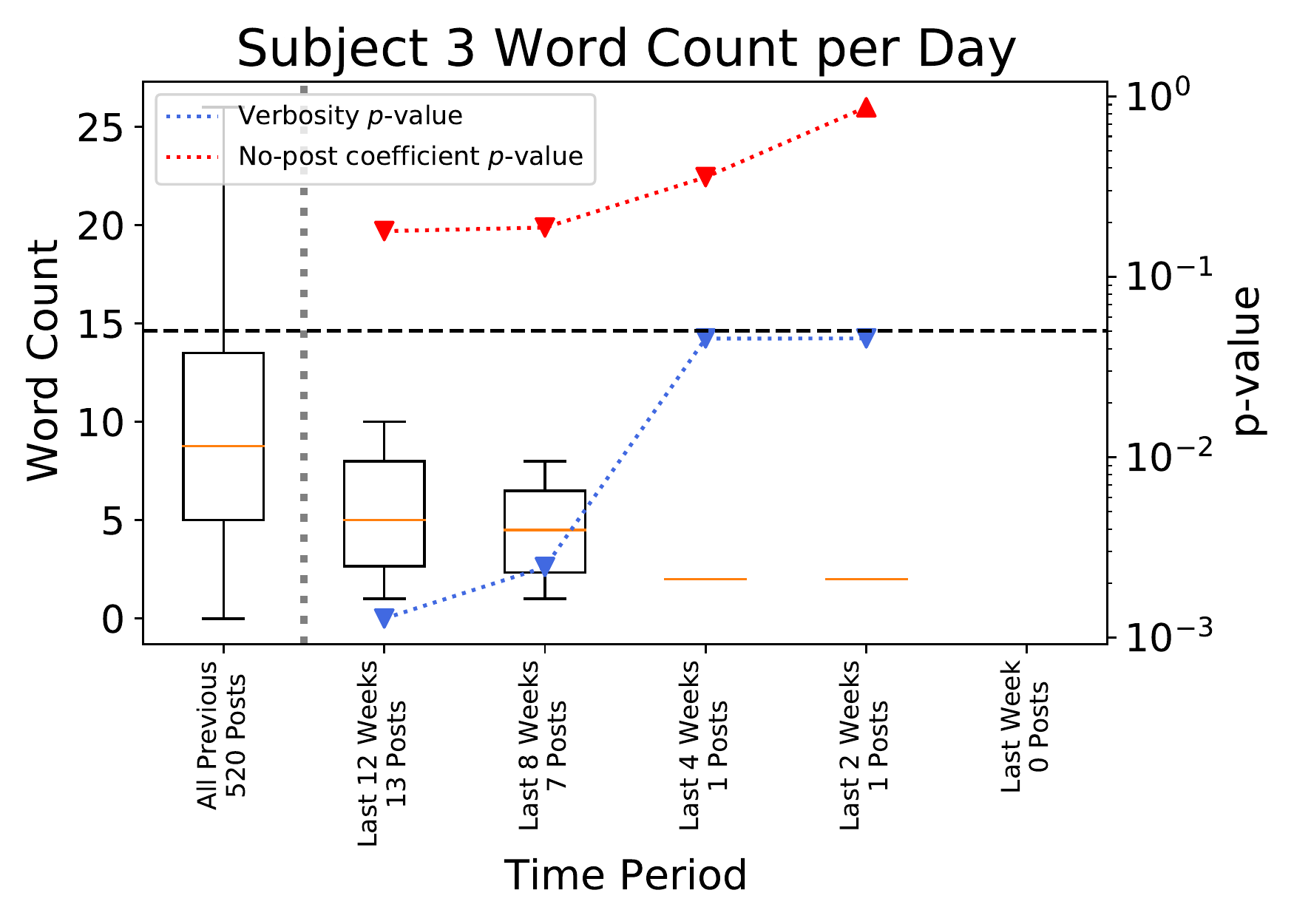}
    \includegraphics[width=.45\textwidth]{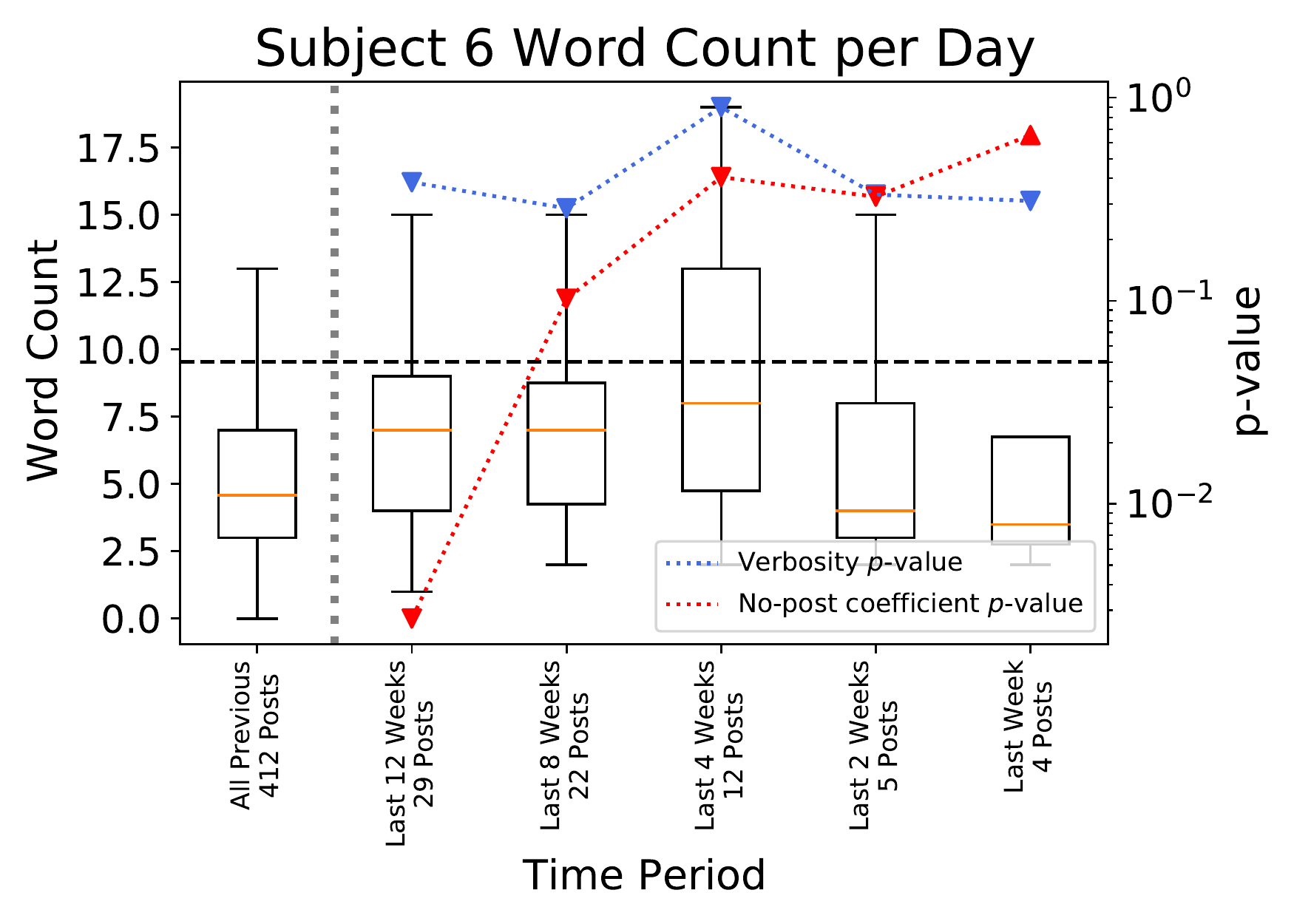}
    \includegraphics[width=.45\textwidth]{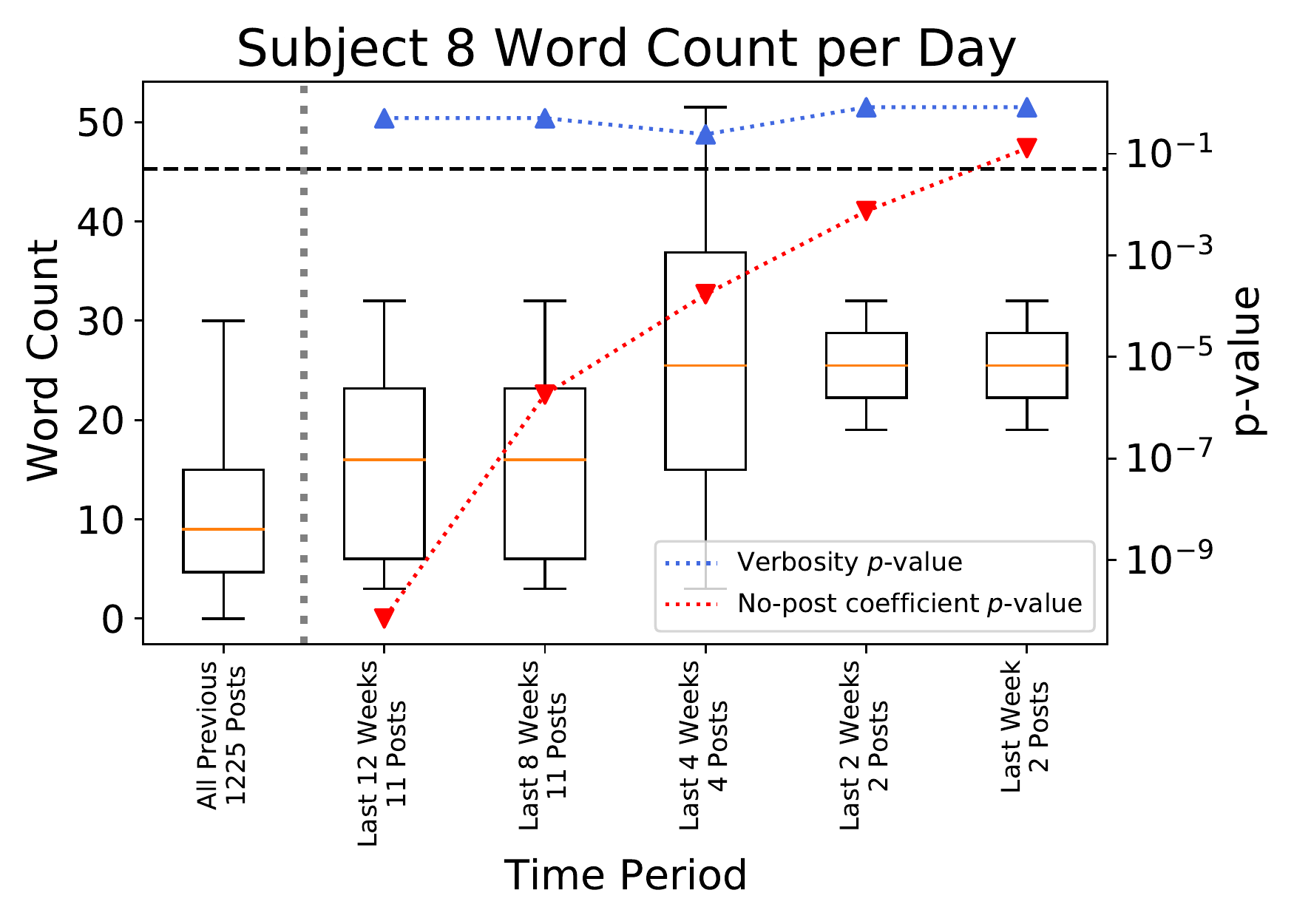}
    \includegraphics[width=.45\textwidth]{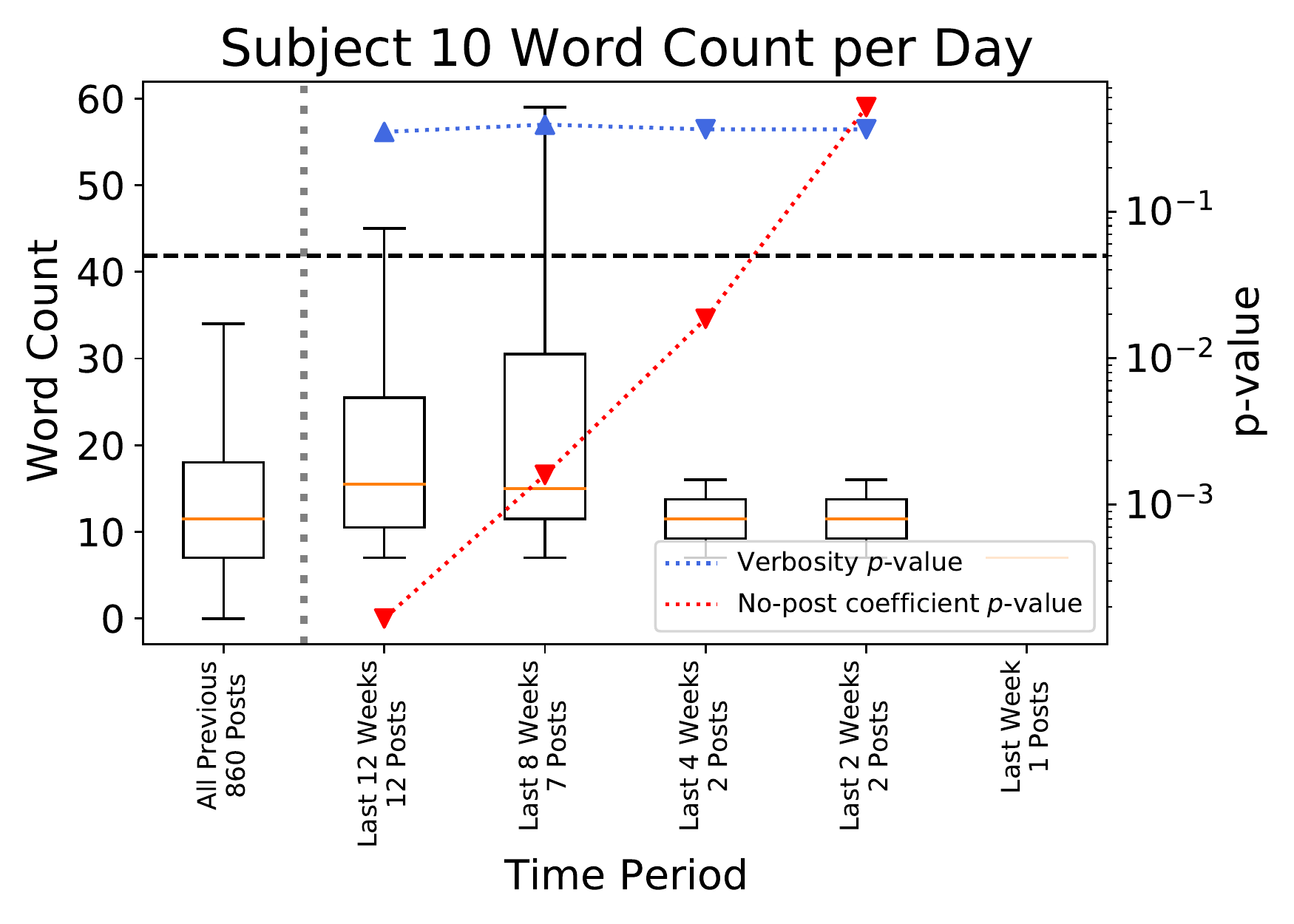}
    \caption{
        \textbf{Subject verbosity per day over different epochs}.
        Difference between word count per day in the period immediately preceding SUDEP compared to word count per day during earlier posting periods.
        Different selections of the time window for the last posting period are displayed on the x-axis.
        The box plot on the far left represents all posts before the 12 weeks preceding SUDEP.
        The blue line represents the p-value of the word count time coefficient for the zero-inflated negative binomial regression.
        The direction of the blue triangle represents the sign of the coefficient, up indicates an increase in wordcount during the period preceding SUDEP and down indicates a decrease.
        The red line represents the p-value of the zero post time coefficient of the regression, with red triangles representing whether there is an increase in the likelihood of any post on a day (up) or a decrease (down).
        The horizontal black line represents $p=0.05$.
    }
    \label{fig:si:wordcount-znb-sigtests-over-time}
\end{figure}

\begin{table}[h!]
    \footnotesize
    \centering
    \begin{tabular}{c|r|r|r|r|r|r|r|r|r|r}
    \toprule
    Subject & $\mu_{1}$ & $n_{1}$ & $\mu_{2}$ & $n_{2}$ & $intercept$ & $time_{coef}$ & $time_{se}$ & $time_{p}$ & $\theta$ & $\theta_{se}$\\
    \midrule
    2 & 12.431 & 2162 & 34.413 & 109 & 2.520 & 1.018 & 0.086 & \textbf{1.197e-32} & 1.373 & 0.043 \\
    1 & 9.592 & 1547 & 17.889 & 54 & 2.261 & 0.623 & 0.135 & \textbf{4.146e-06} & 1.113 & 0.043 \\
    8 & 12.070 & 2185 & 18.375 & 16 & 2.491 & 0.420 & 0.241 & 0.081 & 1.153 & 0.036 \\
    6 & 5.252 & 717 & 7.304 & 23 & 1.659 & 0.330 & 0.143 & \textbf{0.021} & 3.136 & 0.269 \\
    10 & 13.983 & 1147 & 23.571 & 7 & 2.638 & 0.522 & 0.264 & \textbf{0.048} & 2.254 & 0.105 \\
    3 & 11.125 & 834 & 4.100 & 10 & 2.409 & -0.998 & 0.312 & \textbf{0.001} & 1.385 & 0.072 \\
    \bottomrule
    \end{tabular}
    \caption{
        %\rionbr{If we are only analyzig users with more than 500 words, we have to remove other from here. Also, is this table an expanded version of Table \ref{tab:wordcount-sigtests}?} \iwood{Yes, but the main expansion was to include all 12 subjects}
        Statistics from a Negative Binomial Regression on Word Count per Post.
        $\mu_1$ and $n_1$ correspond to the mean word count and number of posts before the last two months, while $\mu_2$ and $n_2$ correspond to the mean word count and number of posts during the last two months before SUDEP. Also included are the $intercept$ of the regression, the coefficient on the last month indicator variable $time_{coef}$, its standard error $time_{se}$, the p-value of the coefficient ${time_p}$, and the dispersion parameter $\theta$ with its standard error $\theta_{se}$.
    }
    \label{tab:wordcount-nbsigtests}
\end{table}

A negative binomial model is often used to model over-dispersed count data, i.e. when the variance is considerably larger than the mean \cite{zeroinflated_examples}. Here a negative binomial model is estimated through a generalized linear regression with log link function on word count per post over a dummy variable representing whether the post's word count occurs during the last month. The significance of the time-indicator dummy variable estimates the significance of the change in the last month over all other posts. As shown in Table \ref{tab:wordcount-nbsigtests} we see significant increases in the word count per post for four subjects at $p < 0.05$. The table is ordered according to the rank product of the number of posts before and during the last two months preceding SUDEP, and the two with the greatest number of posts in both periods by rank product are also the two with the greatest increase in word count, subjects 2 and 1, with two additional subjects showing significant increases, subject 6 and 10. There are five subjects with decreases in word count per post, with subject 11 and subject 3 with significant decreases.

\begin{table}[]
    \footnotesize
    \centering
    \begin{tabular}{c|r|r|r|r|r|r|r}
    \toprule
    Subject & $intercept$ & $time_{coef}$ & $time_{se}$ & $time_{p}$ & $0_{intercept}$ & $0_{time_{coef}}$ & $0_{time_{p}}$ \\
    \midrule
    2 & 3.114 & 1.185 & 0.153 & \textbf{8.802e-15} & -0.275 & -1.854 & \textbf{7.779e-05} \\
    1 & 2.821 & 0.621 & 0.187 & \textbf{8.828e-04} & 0.586 & -0.755 & \textbf{0.007} \\
    8 & 3.028 & 0.213 & 0.318 & 0.503 & -0.261 & 1.637 & \textbf{1.802e-06} \\
    6 & 2.170 & -0.213 & 0.199 & 0.285 & -0.183 & 0.490 & 0.102 \\
    10 & 2.914 & 0.240 & 0.281 & 0.393 & 0.513 & 1.295 & \textbf{0.002} \\
    3 & 2.829 & -1.234 & 0.408 & \textbf{0.002} & 0.991 & 0.571 & 0.187 \\
    \bottomrule
    \end{tabular}
    \caption{
        Statistics of a Zero-Inflated Negative Binomial Regression on word count per day.
        This is similar to Table \ref{tab:wordcount-nbsigtests}, but models the word count per day rather than per post, with the addition of a logistic regression model representing the likelihood of no post at all.
        Included are the $intercept$ of the regression, the coefficient on the last month indicator variable $time_{coef}$, its standard error $time_{se}$, the p-value of the coefficient ${time_p}$.
        Additionally, parameters of the logistic regression on no-post probabilities are shown: the intercept $0_{intercept}$, the coefficient on the time indicator $0_{time_{coef}}$ and the significance of this coefficient $0_{time_{p}}$.
    }
    \label{tab:si:wordcount-znbsigtests}
\end{table}

An alternative formulation is to examine word count per day rather than per post. Perhaps some subjects additionally start posting short posts with increased frequency during periods of stress. However, many days contain zero posts, thus zero words, for most subjects. We can model this with a zero-inflated negative binomial model that also estimates a probability that no words will be posted \cite{zeroinflated_examples, pscl_rlibrary}. As shown in Table \ref{tab:si:wordcount-znbsigtests} we see that subject 2 and 1 still have significant increases in word count per day (columns $time_{coef}$ and $time_{p}$) and both are significantly more likely to post during the last 2 months (columns $0_{time_{coef}}$ and $0_{time_{p}}$, note the negative coefficient corresponds to a lower probability of having no posts on a given day). Subjects 8 and 10 are significantly less likely to post during the last two months. Subject 11, however, is significantly more likely to post during the last two months, although with significantly fewer words per day. Subject 3 is seen to have a significant drop in word count per day. Additionally, subjects 7 and 5 are significantly more likely to post in the last two months, but with non-significant changes in word count. This view of the posting behavior also reveals interesting patterns but is not particularly more informative than the negative binomial model per post.

\end{document}